\pdfoutput=1
\documentclass{aa}
\usepackage{graphicx}
\usepackage{siunitx}
\usepackage[breaklinks]{hyperref}
\usepackage{upgreek}

\DeclareSIUnit\year{yr}
\DeclareSIUnit\parsec{pc}
\DeclareSIUnit\au{AU}
\DeclareSIUnit\magnitude{mag}
\DeclareSIUnit\jansky{Jy}
\DeclareSIUnit\beam{beam}
\DeclareSIUnit\am{arcmin}

\usepackage{url}
\usepackage{txfonts}
\graphicspath{{./figures/}}

\begin{document}

   \title{Large-scale numerical simulations of star formation put to the test}
   
   \subtitle{Comparing synthetic images and actual observations for statistical samples of protostars}

   \author{S. Frimann
          \and
          J. K. J\o rgensen
          \and
          T. Haugb\o lle
          }

   \institute{Centre for Star and Planet Formation, Niels Bohr Institute and Natural History Museum of Denmark, University of Copenhagen, \O ster Voldgade 5-7, DK-1350 Copenhagen K, Denmark \\ \email{sfrimann@nbi.ku.dk} \label{inst1}}

   \date{Received; accepted}
 
  \abstract
   {Both observations and simulations of embedded protostars have progressed rapidly in recent years. Bringing them together is an important step in advancing our knowledge about the earliest phases of star formation.}
   {To compare synthetic continuum images and spectral energy distributions (SEDs), calculated from large-scale numerical simulations, to observational studies, thereby aiding in both the interpretation of the observations and in testing the fidelity of the simulations.}
   {The adaptive mesh refinement code, \texttt{RAMSES}, is used to simulate the evolution of a \SI{5 x 5 x 5}{\parsec} molecular cloud. The simulation has a maximum resolution of \SI{8}{AU}, resolving simultaneously the molecular cloud on parsec scales and individual protostellar systems on AU scales. The simulation is post-processed with the radiative transfer code \texttt{RADMC-3D}, which is used to create synthetic continuum images and SEDs of the protostellar systems. In this way, more than \num{13000} unique radiative transfer models, of a variety of different protostellar systems, are produced.}
   {Over the course of \SI{0.76}{Myr} the simulation forms more than 500 protostars, primarily within two sub-clusters. The synthetic SEDs are used to the calculate evolutionary tracers $T_\mathrm{bol}$ and $L_\mathrm{smm}/L_\mathrm{bol}$. It is shown that, while the observed distributions of the tracers are well matched by the simulation, they generally do a poor job of tracking the protostellar ages. Disks form early in the simulation, with \SI{40}{\percent} of the Class~0 protostars being encircled by one. The flux emission from the simulated disks is found to be, on average, a factor \num{\sim 6} too low relative to real observations; an issue that can be traced back to numerical effects on the smallest scales in the simulation. The simulated distribution of protostellar luminosities spans more than three order of magnitudes, similar to the observed distribution. Cores and protostars are found to be closely associated with one another, with the distance distribution between them being in excellent agreement with observations.}
   {The analysis and statistical comparison of synthetic observations to real ones is established as a powerful tool in the interpretation of observational results. By using a large set of post-processed protostars, which make statistical comparisons to observational surveys possible, this approach goes beyond comparing single objects to isolated models of star-forming cores.}


   \maketitle
%

\newcommand{\LsubLbol}{$L_\mathrm{smm}/L_\mathrm{bol}$\xspace}
\newcommand{\Lbol}{$L_\mathrm{bol}$\xspace}
\newcommand{\Tbol}{$T_\mathrm{bol}$\xspace}
\newcommand{\ramses}{\texttt{RAMSES}\xspace}
\newcommand{\radmc}{\texttt{RADMC-3D}\xspace}
\newcommand{\jzeroeight}{J08\xspace}


\section{Introduction}
The study of star formation is in an era of rapid development. Over the last decade, infrared surveys of nearby molecular clouds, e.g. from the Spitzer Space Telescope and the Herschel Space Observatory, have dramatically increased the number of known young stellar objects (YSOs). Additionally, these surveys have contributed to a better understanding of some of the key questions in star formation, including the evolutionary timescales of YSOs and the distribution of protostellar luminosities (see \citealt{Dunham:2014br} for a recent review). At the same time, sub-millimetre and millimetre interferometers, such as the Submillimeter Array (SMA) and the Atacama Large Millimeter Array, have become able to resolve the small-scale structure, around very deeply embedded objects. Such observations have, for example, demonstrated the presence of Keplerian disks around several Class~I sources \citep[e.g.][]{Brinch:2007gz,Lommen:2008em,Harsono:2014cu}, and around a few Class~0 sources \citep{Tobin:2012ee,Murillo:2013fn,Lindberg:2014bq}.

Increasing computing power has also fuelled significant progress in the field of numerical simulations. For example, a number of studies following the collapse of protostellar cores and the formation of disks and outflows have recently appeared \citep[e.g.][]{Commercon:2012iq,Commercon:2012cm,SantosLima:2012kl,SantosLima:2013hy,Li:2013gu,Myers:2013jv,Seifried:2013hk,Nordlund+2014}. Such studies typically follow the collapse of a single core, resulting in the formation of a single star. Using adaptive mesh refinement (AMR) it is possible to simulate a molecular cloud on parsec scales, while simultaneously resolving the environment around individual protostars on AU scales (\citealt{Padoan:2012jv,Padoan:2014ho}; Haugb{\o}lle et al.\ in preparation). The advantages of this approach are that the influence of the large-scale environment, on the protostellar evolution, is automatically included, and that a global simulation, which forms a large number of protostars, makes it possible to study star formation in the simulation in a statistical manner.

It is an important task to bring together the fields of observations and numerical simulations. Simulations can provide valuable insights into the physics behind observations, while observations are important for validating the simulations. Typically, such validations are done by inferring a physical parameter from the observations, e.g. the initial mass function (IMF), which is then compared to the same parameter in the simulation. Another option is to use forward modelling, and create synthetic observables from the simulations, which can then be compared directly to observations \citep{Padoan:1998hu}. The advantage of the latter approach is that it is generally easier, and involves fewer assumptions, to transform a three-dimensional physical model to synthetic observables, than the other way around. Examples of studies that use synthetic observables to predict observational signatures of different types of YSOs include \citet{Commercon:2012iq,Commercon:2012cm} who predicted the observational signatures of first hydrostatic cores, \citet{Dunham:2012ic} who studied episodic accretion as a solution to the luminosity problem, and \citet{Mairs:2014jn} who looked at the evolution of starless cores in molecular clouds.

This paper presents synthetic continuum images and SEDs\footnote{The synthetic observables and the various calculated parameters presented in the paper have been made available on-line on the web-page: \url{http://starformation.hpc.ku.dk/index.php/synthetic-observations}} of protostars created from a simulation with an unprecedented spatial dynamic range of $2^{17}$ (\num{\approx 130000}:1), encompassing simultaneously molecular cloud and protostellar system scales. The synthetic observables are compared directly with a number of observational studies. The outline of the paper is as follows: Section~\ref{sec:methods} introduces the numerical simulation and the post-processing used to create the synthetic observables. Section~\ref{sec:physicalsimulation} describes the physical characteristics of the simulation, including the identification of cores and disks. Section~\ref{sec:classification} deals with protostellar classification and the ability of observationally defined tracers to follow the physical evolution of protostellar systems. Section~\ref{sec:results} compares the synthetic images and SEDs to three different observational studies: Section~\ref{sec:disk} focuses on disk formation, compared to the continuum survey of \citet{Jorgensen:2009bx}; in Sect.~\ref{sec:luminosities} we compare the protostellar luminosity function (PLF) in the simulation to its observed counterpart \citep{Dunham:2014br}; and in Sect.~\ref{sec:coreprotostar} we study the relationship between protostellar cores and protostars, compared to submm and infrared continuum images from Perseus and Ophiuchus \citep{Jorgensen:2008gz}. Finally, Sect.~\ref{sec:summary} summarises the findings of the paper. 

\section{Methods}
\label{sec:methods}

This section introduces the simulation and the post-processing methods used for creating the synthetic images and SEDs. Only a short description of the technical aspects of the simulation and of the sink particle implementation is presented here, while a more detailed discussion of the sink particle implementation is given in Haugb{\o}lle et al.\ in preparation. Preliminary results from that study were used as guidance for selecting the numerical parameters of this simulation. We also refer to \citet{Padoan:2014ho}, who presented a simulation very similar to the one analysed in this paper.

\subsection{The numerical simulation}
\label{sec:simulation}

The simulation is carried out using the public AMR code \ramses \citep{Teyssier:2002fj}, modified extensively to include random turbulence driving, a novel algorithm for sink particles, and many technical improvements allowing for efficient scaling to thousands of cores. It is one of the largest simulations of a star-forming region ever carried out in terms of number of cells, dynamic range, and number of iterations, and required approximately 15 million CPU hours on the JUQUEEN supercomputer. More than 400 snapshots, with a \SI{5000}{yr} cadence, were stored generating \SI{20}{TB} of raw data and \SI{5}{TB} of post-processed data.

\subsubsection{Initial conditions and physical setup}

We used a finite volume MUSCL scheme with an HLLD method to solve the compressible ideal MHD equations \citep{Teyssier:2006gi,Fromang:2006fi} with an isothermal equation of state in a periodic box.
The model is initialised with a uniform number density of $n_0$\,$\approx$\,\SI{500}{cm^{-3}}, a constant magnetic field strength of $B_0 = \SI{9.4}{\micro G}$, and zero velocity. To create a supersonic turbulent medium, reminiscent of a molecular cloud, we drive the box with a smooth acceleration corresponding to a stirring of the gas. The turbulent driving is done with a solenoidal random forcing in Fourier space at wave numbers $1\le k\le2$ ($k=1$ corresponds to the box-size). A solenoidal force is chosen to guarantee that collapsing regions are naturally generated in the turbulent flow, rather than directly imposed by the driving force. The amplitude is such that the three-dimensional rms sonic Mach number, ${\cal M}_{\rm s}\equiv \sigma_{\rm v,3D}/c_{\rm s}$ (where $\sigma_{\rm v,3D}$ is the three-dimensional rms velocity, and $c_{\rm s}$ is the speed of sound), is kept at an approximate value of 17.

To scale the simulation to physical units, we adopt a temperature $T = \SI{10}{K}$ and a size $L_{\rm box} = \SI{5}{pc}$, which
yields $\sigma_{\rm v,3D}$\,$\approx$\,\SI{3.2}{km.s^{-1}} (consistent with observed line-width size relations), $M_{\rm box}$\,$\approx$\,\num{3670}\,$M_\sun$
(assuming a mean molecular weight of 2.37), and a free-fall time, $t_{\rm ff}$\,$\approx$\,\SI{1.5}{Myr}.  
Gravity is not included during the first 15 dynamical times ($t_{\rm dyn}\equiv L_{\rm box}/ (2 \sigma_{\rm v,3D})\approx \SI{1.5}{Myr}$),
so that the turbulent flow can reach a statistical steady state, and the magnetic energy can be amplified to its saturation level \citep{Federrath+2011}.
Afterwards, the simulation is continued with gravity for a period of \SI{0.77}{Myr} (one dynamical time).
As shown below, this is marginally long enough to allow for the formation of stars of a few solar masses, and thus to sample 
the Salpeter range of the stellar IMF.

\begin{figure*}
\centering
\includegraphics[width=18.6cm]{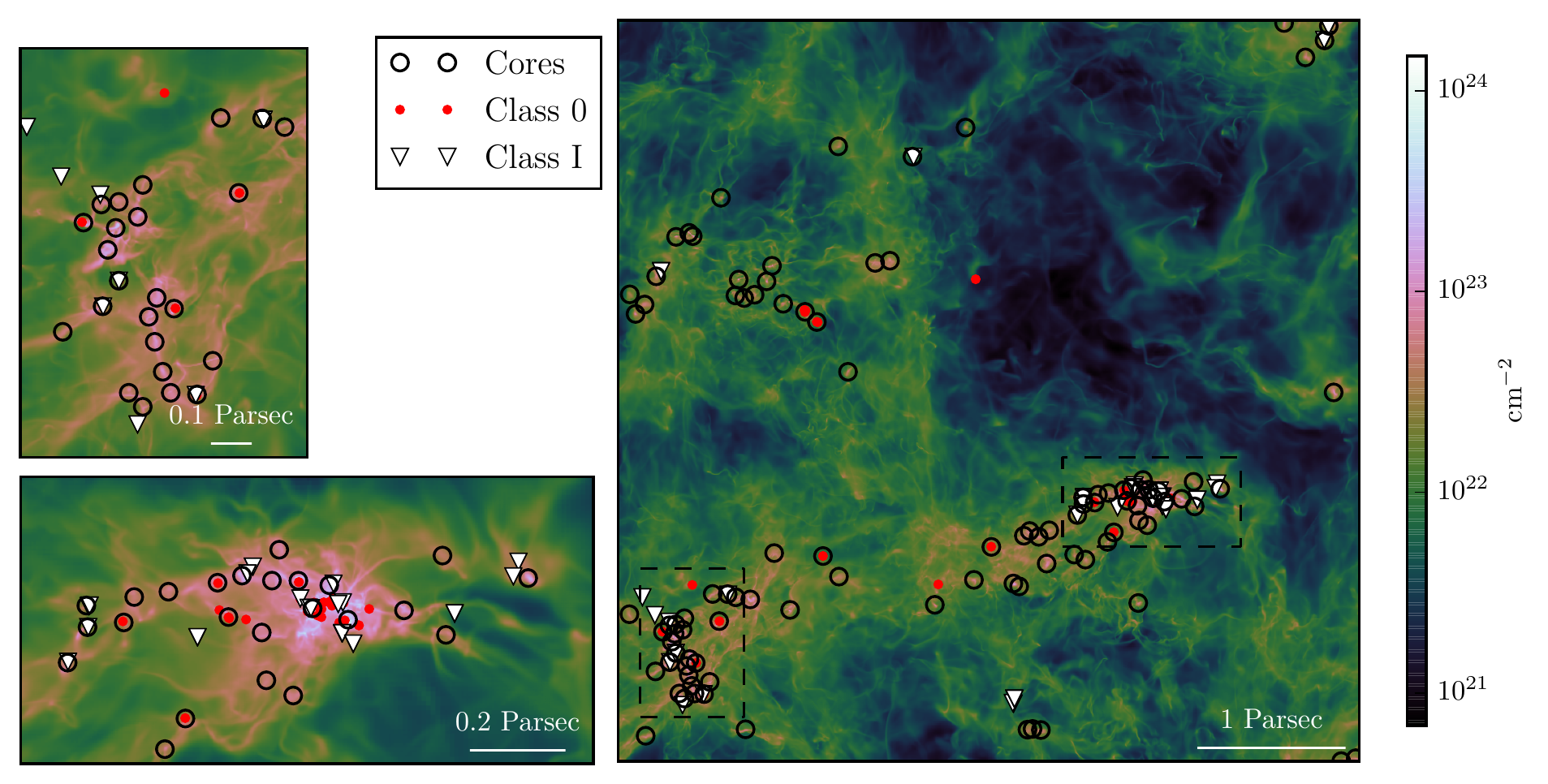}
\caption{Gas column density, protostars, and cores in the simulation \SI{0.6}{Myr} after the formation of the first protostar. The total number of embedded protostars at this point is \num{94}.}
\label{fig:snapshot}
\end{figure*}

The virial parameter, using a practical definition of $\alpha_{\rm vir} \equiv (5/6) \, \sigma_{\rm v,3D}^2 \, L_{\rm box} / (G M_{\rm box})$ \citep{Bertoldi+McKee92}, is 
$\alpha_{\rm vir} = 2.64$. This parameter expresses the ratio between kinetic and gravitational binding energy for a uniform 
isothermal sphere. Its application as an approximate estimate of such energy ratios in simulations is non-trivial, partly because
of the shape and periodic boundary conditions of the numerical box, and partly because of the filamentary distribution of the turbulent gas \citep{Federrath+Klessen2012}.
The high global value of the viral number means that our box corresponds to a loosely bound low-mass star-forming cloud. However, as can be seen in Fig.~\ref{fig:snapshot}, clusters with much lower viral numbers are formed locally in the turbulent flow, with star formation happening predominantly within two sub-clusters inside the box.

\subsubsection{Numerical parameters and the sink particle model}
The root grid of the AMR simulation contains 512$^3$ computational cells with a minimum spatial resolution (in the lowest density regions) of
$\Delta x_{\rm root} = \SI{5}{pc}/512 = \SI{0.01}{pc}$. We use 8~refinement levels, each increasing the spatial resolution by a factor of two. Therefore the maximum spatial resolution
(in dense regions) is $\Delta x = \SI{8}{AU}$. The refinement criterion is based only on density: wherever the density on the root grid, first, and second refinement level are
larger than 8, 64, and 512 times the mean density, one refinement level is added, increasing the resolution by a factor of two. Further levels are added for each increase in
density by a factor of 4, to keep the shortest Jeans length resolved with at least 12 cells at all levels.

A sink particle is created at the highest level of refinement, when the gas density increases above $n \ge n_{\rm max} = \SI{8e9}{cm^{-3}}$ corresponding to $L_{\rm J}\le 6\,\Delta x$.
To create a sink particle it is also required that the gravitational potential has a local minimum in the cell, and that the velocity field is converging, 
$\nabla \cdot {\bf u} < 0$. Furthermore, sink particles cannot be created inside an exclusion radius of $r_{\rm excl}=12\,\Delta x$ around already created sink particles.
These conditions for sink particle creation are similar to those implemented in previous works 
\citep{Padoan:2012jv,Padoan:2014ho,Bate+95sink,Krumholz+04sink,Federrath+10sinks,Gong+Ostriker2013}.

A sink particle is first created without any mass, but is immediately allowed to accrete. In this simulation, it accretes from cells that are closer than an accretion radius of
$r_{\rm acc}=3\,\Delta x= \SI{24}{AU}$, as long as the gas in those cells has a density above a threshold of $n_{\rm acc} = \SI{4e9}{cm^{-3}} = 0.5\,n_{\rm max}$.
Only gas above this threshold is accreted from the cell and onto the sink particle, bringing the gas density slightly below the threshold in the process. The momentum of
the sink particle is changed in accordance with the momentum of the accreted gas, while no magnetic flux is accreted, or removed from the remaining gas. In nature, some
flux is lost due to reconnection and non-ideal effects close to the protostar, on scales smaller than what is reached in this simulation, but this would be non-trivial to include
correctly in a sub-scale model, while maintaining the magnetic field solenoidal.

In nature, YSOs lose a large fraction of their mass due to winds and jets, launched from small scales not included in the simulation. To account for this mass loss, we apply an efficiency factor, $\epsilon_{\rm wind}=0.5$, to all accretion rates and sink particle masses \emph{after} the simulation has finished running. Compared to newer versions of the code, where the mass is removed in situ while running the code \citep{Padoan:2014ho}, and high-resolution zoom-in models around single stars \citep{Nordlund+2014}, this has been shown to be an appropriate value for the resolution used in this simulation.

The characteristic time-step size, of the highest resolution cells in the simulation, is $\Delta t \sim 40$\,days, resulting in roughly \num{7e6} iterations over the \SI{0.77}{Myr} evolution. At the end of the simulation 505 sink particles have been created, containing \SI{3.4}{\percent} of the total initial gas mass. The parameters of the simulation are summarised in Table~\ref{tbl:simulations}.

\begin{figure*}
\centering
\includegraphics[width=18.6cm]{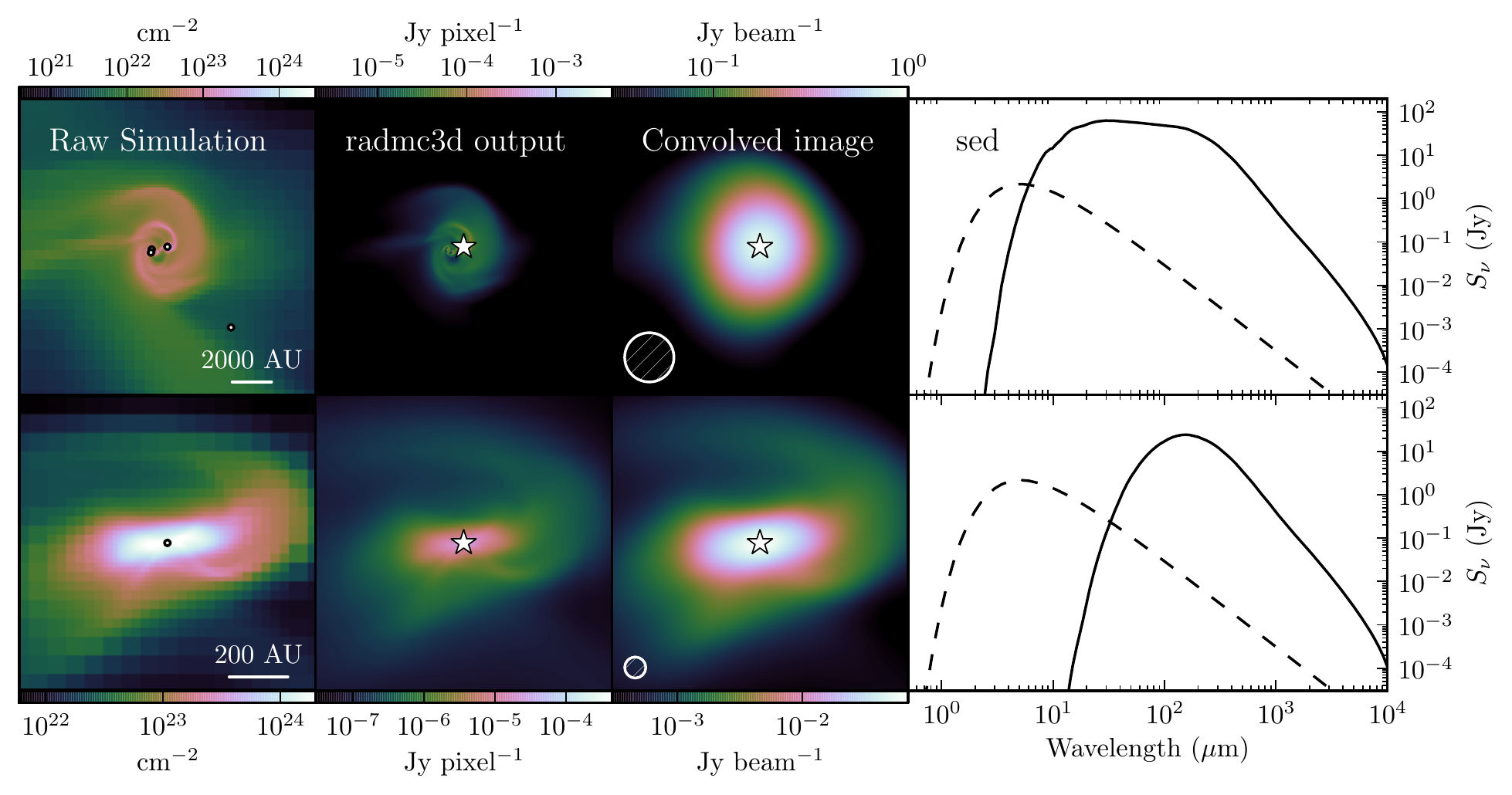}
   \caption{From raw simulation to synthetic observables for two different systems. From left to right: projected gas column density from the raw simulation, with dots indicating protostars; \SI{850}{\micro\metre} \radmc continuum images; continuum images after convolving with a Gaussian beam (Top: \ang{;;15}. Bottom: \ang{;;0.5}); SEDs of the systems, the dashed lines are the SEDs of the central protostars. The assumed distance to both systems is \SI{125}{pc}.}
      \label{fig:continuumimage}
\end{figure*}

To model \emph{ab initio} the formation of individual stars, it is necessary to include much larger scales than those of pre-stellar cores, to avoid imposing ad hoc 
boundary and initial conditions. By driving the turbulence on a scale of \SI{5}{pc}, the formation of cores in the simulation is solely controlled 
by the statistics of the supersonic MHD turbulence, which naturally develops during the initial evolution of 15 dynamical times with no self-gravity. Furthermore,
a box size of \SI{5}{pc} allows the simulation to generate a large number of protostars, sampling well the statistical distribution of  conditions for core formation in the turbulent flow.

The size of the root grid is chosen to be able to resolve the turbulence well everywhere.
The maximum spatial resolution of \SI{8}{AU} is partly dictated by the computational cost of the simulation, and by the goal of following the evolution of a large number of protostars with high enough resolution to resolve their disks in the embedded phase. In the rest of the paper we will refer to the sink particles as ``protostars''.

\subsection{Post-processing}
\label{sec:postprocessing}

The first step in producing synthetic observables from the simulation is to calculate the temperatures of the dust, that are heated by the protostar. To do this, we use the dust radiative transfer code \radmc\footnote{\url{http://www.ita.uni-heidelberg.de/~dullemond/software/radmc-3d/}} (see \citealt{Dullemond:2004iy} for a description of the 2D version of this code). \radmc can handle AMR grids natively, and it is therefore not necessary to resample the density structures from the simulation onto a regular grid.

The total protostellar luminosity, $L_\star$, is modelled as the sum of the accretion luminosity, $L_\mathrm{acc}$, due to mass accretion onto the protostar, and the photospheric luminosity, $L_\mathrm{phot}$, due to deuterium burning and Kelvin-Helmholtz contraction
\begin{equation}
  L_\star = L_\mathrm{acc} + L_\mathrm{phot} = f_\mathrm{acc} \frac{G \dot{m} m_\star}{r_\star} + L_\mathrm{phot}. \nonumber
\end{equation}
Here, $m_\star$ is the mass of the protostar, $\dot{m}$ the accretion rate onto the protostar, $r_\star$ the protostellar radius, and $f_\mathrm{acc}$ is the fraction of accretion energy radiated away. $L_\mathrm{phot}$ is calculated using the pre-main-sequence tracks of \citet{DAntona:1997vs}, where we follow \citet{Young:2005ic} and add \SI{100}{kyr} to the tabulated ages, to account for the time difference between the beginning of core-collapse and the onset of deuterium burning. The accretion rate, $\dot{m}$, is calculated by recording the protostellar mass difference between individual snapshots. The typical snapshot cadence is \SI{\approx 5000}{yr} meaning that the accretion rates are averaged over this time interval. To calculate the accretion luminosity, we assume a stellar radius of $2.5$\,$R_\sun$, while $f_\mathrm{acc}$ is assumed to be 1. All protostars are assumed to emit as perfect black bodies with an effective temperature, $T_\mathrm{eff}$, of \SI{1000}{K}.

The simulation contains more than 500 protostars and 200 million cells, and, because of memory constraints, the dust temperatures cannot be calculated simultaneously in the entire simulation. Instead, \radmc is run on cubical cut-outs centred around individual protostars. These cut-outs have side lengths of \SI{\approx 30000}{AU}, where the exact sizes depend on the arrangement of the AMR levels. The cut-outs are made by cycling through each protostar in each snapshot, with every cut-out corresponding to one \radmc model. Applying this procedure stringently, would yield a total of \num{44531} \radmc models. However, as described below, a number of reductions are made to this sample, bringing the total number of unique \radmc models down to \num{13632}. Each individual cut-out may contain several protostars, but for each \radmc model only one source of luminosity, originating from the central protostar, is included (see Appendix~\ref{sec:multiplestar} for a discussion about how the inclusion of multiple sources of luminosity would affect the results).

\begin{table*}
\caption{Simulation parameters.}
\label{tbl:simulations}
\centering
\begin{tabular}{c c c c c c c c c}     
\hline\hline
$L_\mathrm{box}$ (\si{pc}) & $M_\mathrm{box}$ ($M_\sun$) & $\left\langle n \right\rangle$ (\si{cm^{-3}}) & $\Delta x_\mathrm{min}$ (\si{AU}) &  $\Delta x_\mathrm{max}$ (\si{AU}) & $t$ (\si{Myr})\tablefootmark{a} & $N_\mathrm{snapshot}$ & $N_\mathrm{star}$ & $N_\mathrm{model}$\tablefootmark{b} \\
\hline
5 & \num{3670} & 505 & 8 & 2014 & 0.76 & 188 & 505 & \num{13632}  \\
\hline

\hline                  
\end{tabular}
\tablefoot{
\tablefoottext{a}{Age of oldest protostar at the end of the simulation.}
\tablefoottext{b}{Total number of \radmc models. Note, that because of reductions to the sample, and the fact that not all protostars are present in all snapshots, the total number of models is not equal to $N_\mathrm{star} \times N_\mathrm{snapshot}$ (see Sect.~\ref{sec:postprocessing} for more details).}
}
\end{table*}

Because of the technical set-up of the code, the simulation is not representative for more evolved protostars. The refinement criterion for the AMR levels depends solely on density, which is sufficient to follow the gravitational collapse. However, once the central density in a protostellar system falls below a certain threshold value, the spatial resolution starts dropping as well. A lower resolution leads to an increase of the the numerical viscosity, which in turn increases the rate with which the remaining material close to the protostar is either accreted or dispersed, accelerating the process, and leaving a ``naked'' protostar behind. To follow the protostellar evolution into the less embedded phases, it would be necessary to change the refinement criteria to retain high resolution around the protostars, even when the density in the inner regions start dropping. For late evolutionary stages the radiation of the central protostar on the physical structure (in particular, in the circumstellar disk) also becomes increasingly important. Consequently, in the following analysis, we will only use the embedded objects, and require the environments around the protostars to be as well resolved as possible: to include a system in the analysis, we require that the protostar lie in an AMR cell of level 5 or higher, corresponding to a cell size <\,\SI{63}{AU} and a minimum number density of \SI{4e6}{cm^{-3}}.

Some protostars are too faint to be detected by infrared surveys, like the Cores to Disks (c2d) Spitzer survey \citep{Evans:2009bk} and similar. Such survey are typically complete down to a luminosity of \num{\sim 0.05}\,$L_\sun$. In the simulation, we therefore assume that all protostars with a luminosity below this value are too faint to be detected, and they are removed from the sample.

We assume a uniform dust-to-gas mass ratio of 1:100 everywhere, and make use of the dust grain opacities of \citet{Ossenkopf:1994tq}, corresponding to coagulated dust grains with thin ice mantles at a density of $n_\mathrm{H2}$\,$\sim$\,\SI{e6}{cm^{-3}}. These opacities have been found, by several studies \citep[e.g.][]{vanderTak:1999gi,Shirley:2002co,Shirley:2011jz}, to be appropriate for dense cores. The opacities do not extend beyond \SI{1.3}{mm}, and are therefore extrapolated at longer wavelengths, using a power law ($\kappa_\nu$\,$\propto$\,$\nu^\beta$ with $\beta$\,=\,1.7). \radmc takes absorptive dust opacities, $\kappa_\mathrm{abs}$, as input while the opacities tabulated in \citeauthor{Ossenkopf:1994tq} are total ones, including both scattering and absorption, $\kappa_\mathrm{tot} = \kappa_\mathrm{abs} + \kappa_\mathrm{scat}$. \citet{Dunham:2010bx} demonstrated that $\kappa_\mathrm{scat}$ dominates over $\kappa_\mathrm{abs}$ between \SIlist{0.1;10}{\micro\metre}. This study is mainly concerned with longer wavelengths, where scattering can be safely ignored.

\radmc uses the Monte Carlo method of \citet{Bjorkman:2001du} to calculate the dust temperatures. This method relies on the propagation of a number of ``photon packets'' through the model, which, in our case, has been set to one million. The optically thin parts of the resulting temperature profiles roughly follow a power law, $T_\mathrm{dust} \propto r^{-\beta}$ with $\beta$\,$\approx$\,0.4.

Once the dust temperatures have been calculated, \radmc is used to calculate continuum images and SEDs (see Fig.~\ref{fig:continuumimage} for two examples). The continuum images are subsequently convolved with a Gaussian beam to simulate single-dish observations, or sampled in the $(u,v)$-plane to simulate interferometric observations. The SEDs are calculated by integrating the emission over a square aperture with side lengths of \SI{2250}{AU}, corresponding to \ang{;;15} at \SI{150}{pc}, centred around the central object. As standard, three orthogonal directions are sampled when calculating continuum images and SEDs, effectively increasing the amount of data with a factor of three.

\section{Physical description of simulation}
\label{sec:physicalsimulation}

\begin{figure}
\includegraphics[width=\hsize]{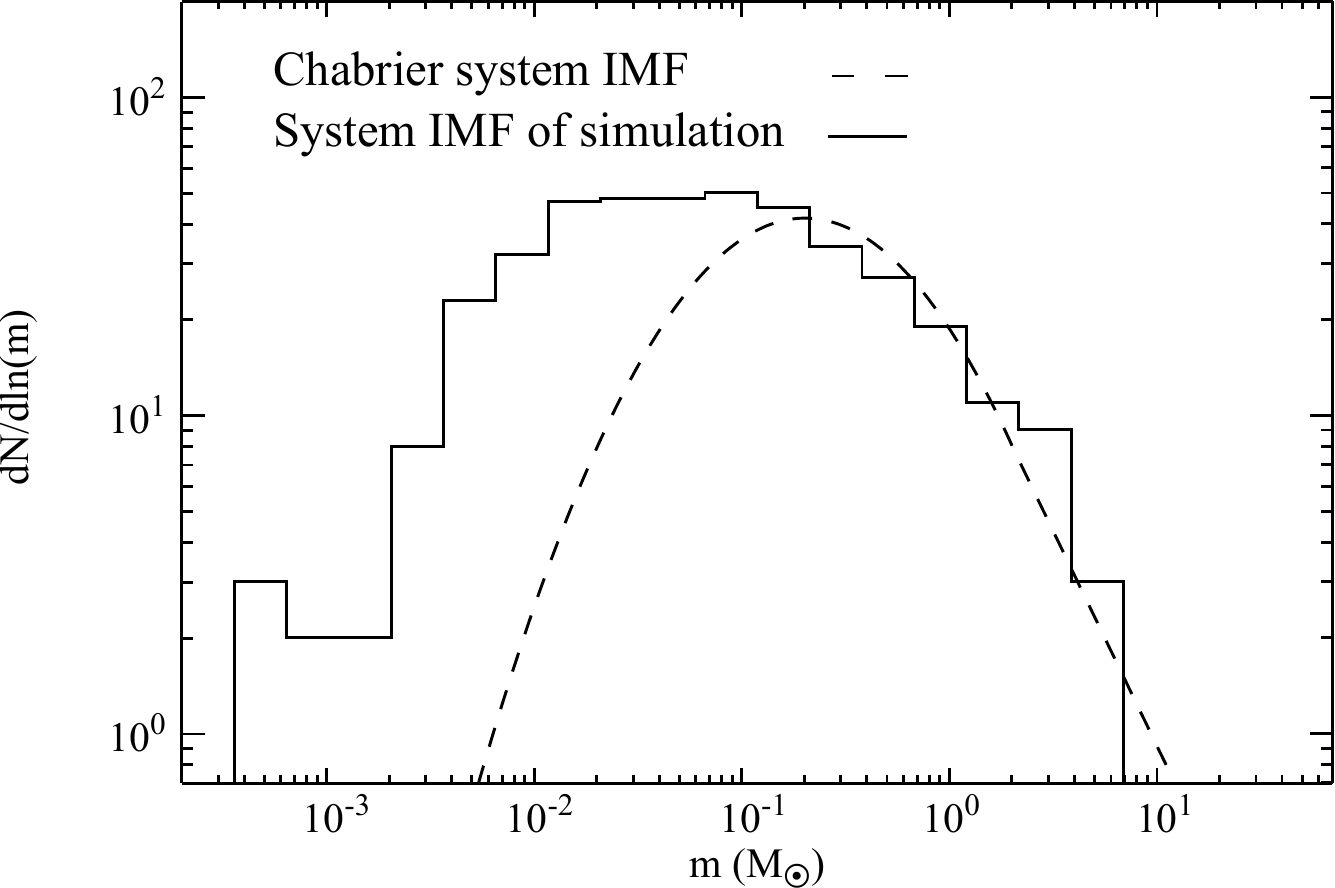}
\caption{Initial mass function in the simulation containing \num{\sim 500} protostars distributed in 429 systems and sampled \SI{0.76}{Myr} after the formation of the first protostar. The dashed line is the corresponding Chabrier system IMF. We have excluded stars, which either have an accretion rate above \num{1e-5}\,$M_\sun$\,yr$^{-1}$ (2 stars), would double their mass in less than \SI{100}{kyr} (9 stars), or stars which are younger than \SI{50}{\kilo\year} (23 stars). This removes objects, which would not normally be included in an IMF, because they are heavily embedded, either due to young age or high accretion rates.}
\label{fig:imf}
\end{figure}

\subsection{General overview}
\label{sec:generaloverview}

Table~\ref{tbl:simulations} summarises key parameters of the simulation. The mean gas density in the simulation is within a factor of two of several nearby molecular clouds, such as Cha~II, Lupus, and Ophiuchus \citep{Evans:2009bk}. Of these, the simulation resembles Ophiuchus, which is still actively forming stars, the most. Approximately $80 \, M_\sun$, or \SI{2}{\percent}, of the gas in the cloud is found to lie at column densities <\,\SI{2e21}{cm^{-2}}, corresponding to a visual extinction threshold $A_V < \SI{2}{mag}$ \citep{Bohlin:1978dw}. This is the same threshold used by \citet{Evans:2009bk} to determine the masses of the clouds in the c2d Spitzer survey. The exact value depends somewhat on the orientation of the cloud relative to the observer, the assumed resolution of the extinction maps, and the age of the cloud, but in any case the majority of the material in the simulation is found in regions with $A_V \geq \SI{2}{mag}$.

Our molecular cloud simulation reproduces well the Salpeter slope of the IMF, except for a clear overproduction of brown dwarfs, relative to the Chabrier system IMF \citep{Chabrier:2003aa} (see Fig.~\ref{fig:imf}). We have made a hierarchical multiplicity analysis, which shows that, at the end of the simulation, the protostars are distributed in 389 single star systems; 56 stars in 28 binaries; 18 stars in 6 triple systems, and 42 stars in 6 multiple systems, including two systems with 11 and 13 members. The median seperation in the binary systems is \SI{150}{AU}, which is higher than what is found observationally. This is a consequence of the \SI{8}{AU} cell resolution, and the \SI{96}{AU} exclusion radius, which preclude the possibility of modelling binaries resulting from disk fragmentation. Most of the brown dwarfs are formed in two dense sub-clusters, and are dynamically expelled at a young age. This brown dwarf population does not affect the conclusions in the paper: most of them are not included in the post-processing, due to low accretion rates, and because of their low mass, they do not affect the mass reservoir available for the rest of the protostars.

Figure~\ref{fig:snapshot} shows the gas column density and positions of the embedded protostars and cores in the simulation \SI{0.6}{Myr} after the formation of the first protostar. The total number of cores and protostars identified in the figure are \num{102} and \num{94} respectively. The number of protostars differs from the \num{505} listed in Table~\ref{tbl:simulations} because not all protostars have formed at this point, and because some protostars have been removed from the sample as described in Sect.~\ref{sec:postprocessing}.

Looking at Fig.~\ref{fig:snapshot}, it is clear that protostars and cores tend to cluster around regions of high column density; something which is also observed in nature (e.g. \citealt{Evans:2009bk}). Most of the star formation in the simulation is situated in two sub-clusters, roughly identified by the rectangular inserts in Fig.~\ref{fig:snapshot}. The two clusters, defined by the inserts, both have a cross sectional area of \SI{0.7}{pc^2} and roughly the same mass ($285 \, M_\sun$ and $276 \, M_\sun$ respectively). The more massive cluster hosts 57 embedded protostars and 24 cores, while the less massive one hosts 15 embedded protostars and 24 cores. The two clusters began forming stars at roughly the same time, so this discrepancy is not due to time differences in the onset of star formation. An alternative explanation for the discrepancy is variations in the local density, and therefore virial numbers, reflected in the local star formation rates \citep{Padoan:2012jv}. The gas in the less massive cluster is more dispersed than the gas in the massive cluster, which is concentrated around a very dense central region where the majority of the Class~0 protostars are located. More quantitatively, \SI{\sim 20}{\percent} of the gas mass in the massive cluster lie at column densities >\SI{1e23}{cm^{-2}}, while the same is only true for \SI{\sim 2}{\percent} in the less massive cluster. In total, \SI{77}{\percent} of the embedded protostars, and \SI{47}{\percent} of the cores, shown in Fig.~\ref{fig:snapshot}, are located in one of the two clusters.

\subsection{Cores in the simulation}
\label{sec:cores}

Dense cores are the smallest units in the hierarchical structure that molecular clouds are made up of, and can be defined as over-dense regions in the molecular cloud, corresponding to local minima of the gravitational potential. Typically, one distinguishes between protostellar and starless cores, depending on whether they are associated with a protostar or not. Observationally, dense cores are most readily detected by their continuum emission in the submm wavelength range. To this end, we have created synthetic \SI{850}{\micro\metre} continuum images of all the protostars in the simulation, which are used to identify and characterise the cores in the simulation. Because of the finite sizes of the cut-outs, this method misses starless cores lying at distances $\gtrsim$\,\SI{15000}{AU} from their nearest protostar. At \SI{850}{\micro\metre}, the dust can be expected to be optically thin, and we can therefore include the missing regions by converting the raw column density maps, from the simulation, into dust continuum images using the formula, $S_\nu = N\kappa_\nu B_\nu (T_d)$. Here, $N$ is the dust column density, $\kappa_\nu$ the dust opacity, and $B_\nu (T_d)$ the Planck function at a dust temperature, $T_d$, which we assume to take a value of \SI{10}{K}. The continuum images are convolved with a Gaussian beam with FWHM of \ang{;;18}, assuming a distance to the cloud of \SI{250}{pc}. The cores are identified and characterised using the core-finding algorithm, \texttt{CLFIND2D} \citep{Williams:1994hl}. To accept something as a core, we require its peak flux to be \SI{> 0.15}{Jy.beam^{-1}}, and that it is resolved (radius > \ang{;;9}). These choices were made to match the methodology used in the observational studies of \citet{Kirk:2006kt,Jorgensen:2008gz}, to which the core list is compared in Sect.~\ref{sec:coreprotostar}.

There are significant overlaps between the continuum images used for detecting the cores, and hence also a risk of counting individual cores several times. This is solved by checking the final list of cores for overlaps. In case two or more cores overlap, only the core with the highest peak flux is kept, while the rest are discarded. On average, we find \num{\approx 100} cores per snapshot, when assuming a cloud distance of \SI{250}{pc}. Heating from the protostars increase the submm emission, and thereby also the number of core detections, meaning that this number depends on the number of protostars to a large degree, and on the assumed cloud projection and overall cloud evolution to a smaller degree. The earliest snapshots, with only a handful of protostars, contain \num{\sim 60} cores, while, onwards of \SI{0.3}{\mega\year} after the formation of the first protostar, \num{\sim 110} cores per snapshot are detected.

Studies of submm emission and extinction maps of molecular clouds suggest the existence of an extinction threshold, or equivalently a column density threshold, for core formation at $A_V$\,$\sim$\,8 \citep{Johnstone:2004ix,Enoch:2007dc,Konyves:2013gz}. Such a threshold is predicted theoretically by \citet{McKee:1989ho}, whose model of photoionisation regulated star formation prohibits core collapse at extinctions $A_V \lesssim 4-8$. An alternative explanation for the observed extinction threshold is that cores are a product of Jeans fragmentation, and therefore primarily appear in the densest regions of the cloud \citep{Larson:1985to}.

The simulation is isothermal MHD, and does not include any ionising radiation, so any production of an extinction thres\-hold, similar to that seen in observations, cannot be explained by the presence of photoionising radiation. To test for a possible extinction threshold of the cores in the simulation, we first convert the raw column density maps of the full simulated box into visual extinction maps using the conversion $\langle N_{\mathrm{H}_2} \rangle/A_V$\,$\approx$\,\SI{1e21}{cm^{-2}.mag^{-1}} \citep{Bohlin:1978dw}. By comparing the extinction maps to the positions of the cores, identified by their submm emission as described above, the visual extinction of all cores can be calculated and compared to observations.

\begin{figure}
\includegraphics[width=\hsize]{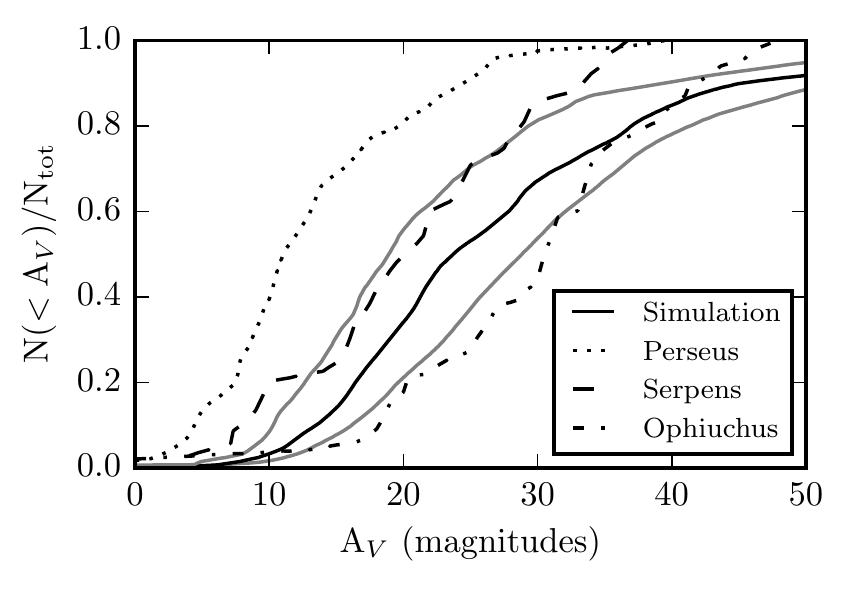}
\caption{Cumulative distributions of visual extinctions of cores in the simulation, along with observations of Perseus, Serpens, and Ophiuchus from \citet{Enoch:2007dc}. The black solid line the simulated extinction curve, assuming a resolution of \ang{;;90} and a distance of \SI{250}{pc}. The grey solid lines show the effect of increasing/decreasing the resolution by a factor of four.}
\label{fig:extinction}
\end{figure}

The black solid line in Fig.~\ref{fig:extinction} shows the cumulative distribution of core extinctions in the simulation, which clearly reproduces an extinction threshold like that seen in observations. Observed distributions from Perseus, Serpens and Ophiuchus, taken from \citet{Enoch:2007dc}, are plotted along with the synthetic data for comparison. The extinction maps, used by \citet{Enoch:2007dc}, have a resolution of \ang{;;90} so the synthetic extinction maps have been down-sampled by running a two-dimensional median filter across them with the same resolution, while assuming a distance to the cloud of \SI{250}{pc}. The grey lines show the effect of increasing/decreasing the resolution of the extinction maps by a factor of four, and has the effect of shifting the distribution \SI{\approx 5}{mag} towards higher extinctions when increasing the resolution, and vice versa when decreasing the resolution.

A glance at Fig.~\ref{fig:extinction} reveals that the simulated distribution lies between the Serpens and Ophiuchus distributions. The shape of the simulated distribution is similar to the observations, with the notable difference that the simulated distribution has a tail towards high extinctions, not seen in any of the observed clouds. This tail is a result of the very dense cluster described at the end of Sect.~\ref{sec:generaloverview}, which likely has no counterpart in any of the observed local star-forming regions. An alternative explanation is that background stars cannot normally be detected through the densest regions of the observed clouds, meaning that the highest column densities might be missed in the observed extinction maps.

There are distinct differences between the observed cumulative distributions of Perseus (distance of \SI{250}{pc}, \citealt{Enoch:2006gf}), Serpens (\SI{415}{pc}, \citealt{Dzib:2010hi}), and Ophiuchus (\SI{125}{pc}, \citealt{deGeus:1989vn}). This is not simply a question of distance as Serpens is furthest away of the three clouds, while its cumulative distribution is intermediate between the other two. As shown in Fig.~\ref{fig:extinction}, changing the resolution by a factor of four (equivalent to changing the assumed distance by a factor of two), is sufficient to explain the difference between the Serpens and Ophiuchus distributions, which are matched well by the two grey lines. The fact that the Perseus distribution is not reproduced indicate that environmental factors also play a role. A detailed study into the nature of these environmental differences, and how they relate to the simulation, is beyond the scope of this work.

\begin{figure*}
\centering
\includegraphics[width=18.6cm]{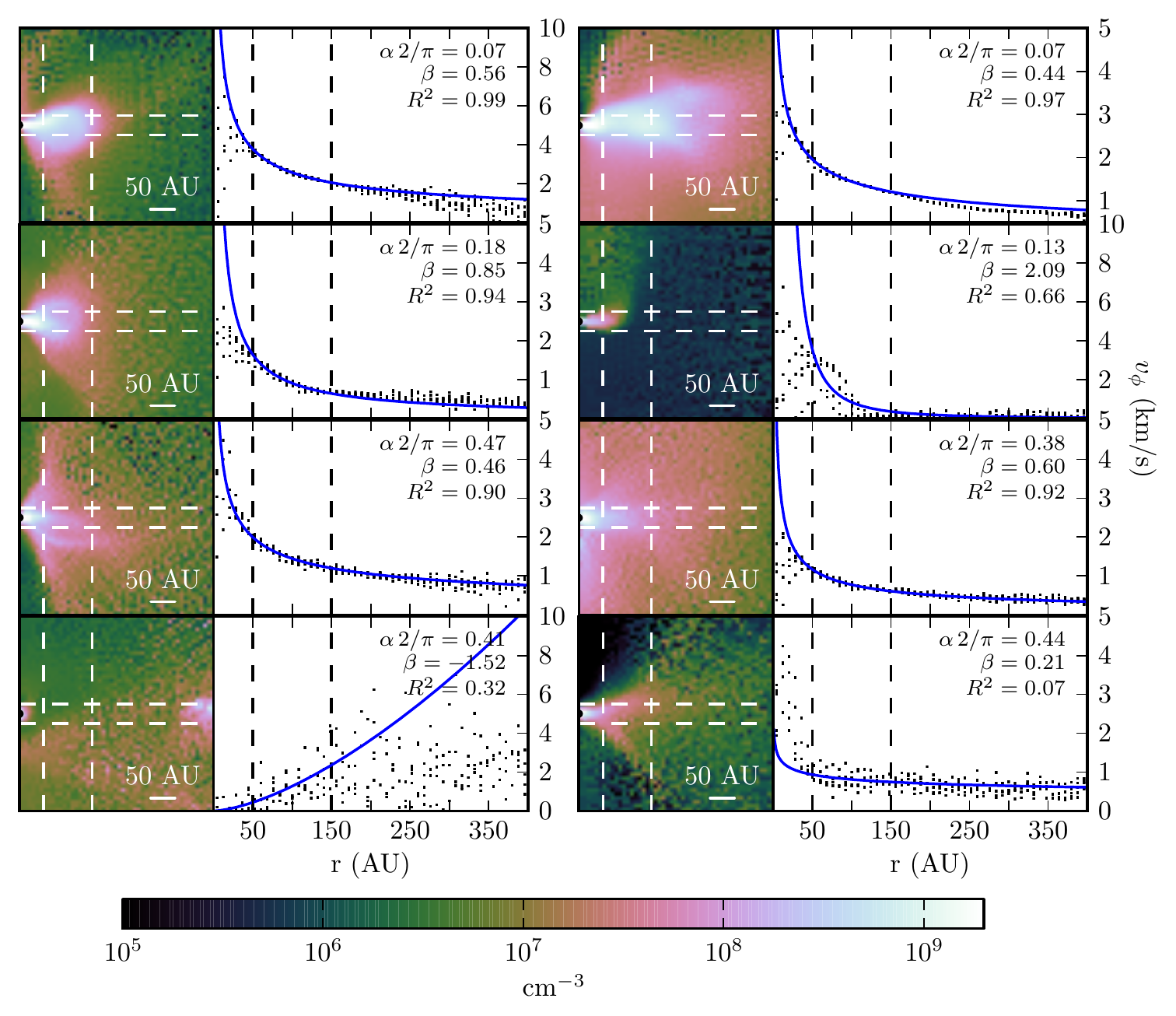}
   \caption{Examples of disks in the simulation. The dashed lines indicate the regions used for fitting the rotation curves for the power law method. The blue solid line in each panel is the best fitting power law. In the top right corner of each panel is given the angle $\alpha$, $R^2$, and the power law index, $\beta$. Top row: systems that are disks by both the $\alpha$-angle and the power law method. Second row: disks by the $\alpha$-angle method, but not by the power law method. Third row: disks by the power law method, but not by the $\alpha$-angle method. Bottom row: disks by neither method.}
      \label{fig:disk}
\end{figure*}

\subsection{Disks in the simulation}
\label{sec:diskinsimulations}

The question of disk formation is a central one for observers and modellers alike. Rotationally supported, or \emph{Keplerian}, disks form as a consequence of angular momentum conservation, and are characterised by their rotational velocity profile, $v_{\phi}(r) \propto r^{-0.5}$. Magnetically supported \emph{pseudodisks} may develop in magnetised cores prior to the formation of Keplerian disks, as magnetic pinching forces deflect infalling material away from the radial direction and towards the disk's midplane \citep{Galli:1993ix,Galli:1993dl}. Magnetically supported pseudodisks are not expected to have the same rotation profiles as Keplerian disks; instead one may assume that the infalling material, in such disks, conserve specific angular momentum, which suggests $v_{\phi}(r) \propto r^{-1}$ \citep{Belloche:2013fq}.

\subsubsection{Disk detection methods}

With a sample consisting of more than \num{13000} individual objects, a simple, yet robust, method for determining if a given system contains a disk, is needed to be able to draw conclusions based on statistics. We have devised a method for this, which we will call the $\alpha$-angle method, based on the ratio between the radial and rotational motions in a system. The first step is to find the rotation axis, $\vec{\Omega}$, of the system.
\begin{equation}
  \vec{\Omega} = \frac{ \sum_i \rho_i \cdot \vec{r}_i \times \vec{v}_i }{\sum_i \rho_i}, \nonumber
  \label{eq:rotationaxis}
\end{equation}
where the sum is over all cells located within a radius of \SI{400}{AU} from the protostar. $\rho_i$, $\mathbf{r_i}$ and $\mathbf{v_i}$ are respectively density, position and velocity of individual cells relative to the central protostar. The result is weighted with density to make the disk's high-density midplane count more towards the determination of the rotation axis. We find that this gives a robust determination of the disk plane, except for the cases where the central protostar is associated with one or more companion protostars, which may interfere with the determination of the rotation axis. Approximately \SI{40}{\percent} of the protostars, in the sample, are accompanied by one or more protostellar companions within the \SI{400}{AU} inclusion radius. In the following, these systems are disregarded.

Using the rotation axis as a reference, the velocities can be resolved into radial and rotational components, $v_{r}$ and $v_{\phi}$, where $v_{r}$ is defined such that the outward direction is positive. We go on to calculate the mass weighted averages of these velocities, $\langle v_{r} \rangle$ and $\langle v_{\phi} \rangle$, within a radius of \SI{400}{AU} from the protostar. The final result is the angle, $\alpha$, defined as
\begin{equation}
 \alpha \equiv \arctan \frac{-\langle v_r \rangle}{\langle v_\phi \rangle}. \nonumber
\end{equation}
$\alpha$ is the angle between the ``average'' velocity vector of the gas in the system and the $\boldsymbol{\hat{\phi}}$ unit vector. $\alpha = \pi/2$ thus corresponds to a system of purely infalling material, $\alpha = 0$ to pure rotation, and $\alpha = -\pi/2$ to pure outflow. This way of characterising the relationship between radial and rotational motions was first introduced by \citet{Brinch:2007jj}, and first used in the context of a numerical simulations by \citet{Brinch:2008cz}. By manually inspecting a large number of systems, we determine that disks are mainly found for values of $\alpha$ lying in the range, $0 \leq \alpha \leq 0.2 \, \pi/2$, which is the criterion adopted for claiming the presence of a disk using this method.

The $\alpha$-angle method is not able to distinguish between Keplerian disks and other types of rotating structures. In an effort to test the performance of the $\alpha$-angle method, we also fit power law functions of the form, $v_\phi(r) = A \, r^{-\beta}$, to the systems in the sample. To fit a power law function, the rotational velocities are averaged in the azimuthal direction, and collected into \SI{10 x 10}{AU} sized bins to create a cross-section of the disk, similar to the density cross-sections shown in Fig.~\ref{fig:disk}. The fitting range is restricted to \SI{20}{AU} above and below the disk's midplane, and between \SI{50}{AU} and \SI{150}{AU} in the radial direction. The inner boundary of \SI{50}{AU} is chosen to avoid issues related to the spatial resolution of the simulation, while the outer boundary of \SI{150}{AU} is chosen to avoid fitting beyond the outer edge of as many disks as possible. As a goodness-of-fit parameter we use $R^2$, which is defined as
\begin{equation}
  R^2 \equiv 1 - \frac{SS_\mathrm{res}}{SS_\mathrm{tot}} = 1 - \frac{\sum_i \left( v_{\phi,i} - f_i \right)^2 }{\sum_i \left( v_{\phi,i} - \langle v_{\phi} \rangle \right)^2}, \nonumber
\end{equation}
where $SS_\mathrm{res}$ and $SS_\mathrm{tot}$ are the squared sum of the residuals, relative to the fit and the mean value respectively. To claim the presence of a Keplerian disk, using the power law method, we require two criteria to be fulfilled: first we ensure that the power law function is a good match to the rotational velocities by requiring that $R^2 > 0.8$; second we require the power law exponent, $\beta$, to fall within the range $0.25 \leq \beta \leq 0.75$.

\subsubsection{Comparing methods of disk detection}

Figure~\ref{fig:disk} shows cross sections of some of the systems in the simulation along with scatter plots of the rotational velocities in the disk's midplane. The top row of panels in Fig.~\ref{fig:disk} shows two systems, for which both the $\alpha$-angle and power law method predict the presence of a disk. The second row of panels shows systems, that are disks by the $\alpha$-angle method, but not by the power law method. The system in the left panel is rejected because the power law exponent, $\beta$, is too steep, while the system in the right panel appears to harbour a \SI{\approx 100}{AU} sized disk, which is too small to be fitted well by the power law method. The third row of panels shows systems, which are disks by the power law method, but not by the $\alpha$-angle method. In both these systems it is difficult to identify any structure, recognisable as a disk, from the density cross-sections. The bottom row of panels shows examples of systems that are disks by neither method. Table~\ref{tbl:disk} presents a quantitative comparison between the $\alpha$-angle and power law methods. The two methods agree \SI{82}{\percent} of the time, and both methods find disks around approximately half the protostars in the sample.

\begin{table}
\caption{Comparison between the $\alpha$-angle and the power law methods for disk detection. To claim detection of a disk, using the $\alpha$-angle method, we require that $0 \leq \alpha \leq 0.2 \, \pi/2$. To claim detection of a Keplerian disk, on scales between \SI{50}{AU} and \SI{150}{AU}, using the power law method, we require the power law exponent to fall into the range $0.25 \leq \beta \leq 0.75$ and $R^2 > 0.8$.}
\label{tbl:disk}
\centering
\begin{tabular}{l l c c c }
\hline\hline
        & & \multicolumn{2}{c}{$\alpha$-angle method} & \\
        & & Disk & No disk & \\
\hline
Power law & Disk    & \SI{42}{\percent} & \SI{7}{\percent}  & \textbf{\SI[detect-weight]{49}{\percent}} \\
method    & No disk & \SI{11}{\percent} & \SI{40}{\percent} & \textbf{\SI[detect-weight]{51}{\percent}}  \\
          &         & \textbf{\SI[detect-weight]{53}{\percent}} & \textbf{\SI[detect-weight]{47}{\percent}} & \\
\hline
\end{tabular}
\end{table}

\begin{figure*}
\sidecaption
\includegraphics[width=12cm]{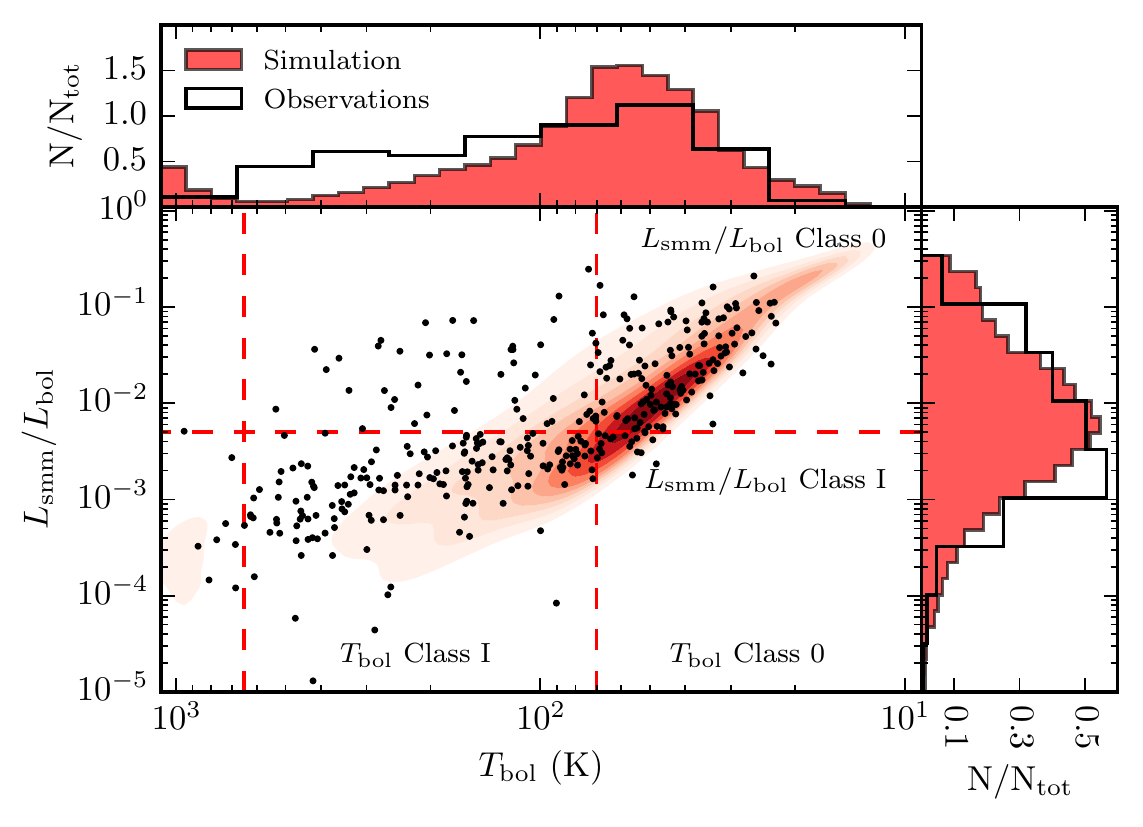}
   \caption{Distribution of \Tbol and \LsubLbol, for the protostars in the simulation (red contours), and from observations (black dots). The contour levels cover \SI{90}{\percent}, \SI{80}{\percent}, \SI{70}{\percent}, {\ldots} of the simulated points. The marginal distributions of each variable are shown as histograms at the edges. The observations are a conjunction of data from c2d, GB, and HOPS (see \citealt{Dunham:2014br}). Protostars are expected to evolve from upper right to lower left. For the fraction of (synthetic) points recorded in each quadrant see Table~\ref{tbl:classclass}.}
   \label{fig:TbolLsub}
\end{figure*}

There is a degree of arbitrariness to the range of $\beta$ values used by the power law method to detect Keplerian disks. Narrowing the range to $0.4 \leq \beta \leq 0.6$ decreases the disk fraction from \SI{49}{\percent} to \SI{35}{\percent}. Thus, even when imposing a more conservative range on the power law exponent, more than one third of the systems are still found to have rotation curves consistent with Keplerian rotation. The results presented in this section illustrate that the $\alpha$-angle method is a robust way of detecting disks in the simulation. Even though the method is not sensitive to the shape of the rotation profile, most of the disks found using this method are consistent with Keplerian rotation. For the remainder of this paper we use the $\alpha$-angle method for disk detection, and claim the presence of a disk if $\alpha$ falls into the range $0 \leq \alpha \leq 0.2 \, \pi/2$.

\section{Classification of protostars}
\label{sec:classification}

Traditionally, YSOs are sorted into four different observationally defined classes: 0, I, II, and III \citep{Lada:1984hg,Lada:1987vy,Andre:1993cz}. Two widely used tracers for determining the class are the bolometric temperature, \Tbol \citep{Myers:1993en}, and the ratio between sub-millimetre and bolometric luminosity, \LsubLbol \citep{Andre:1993cz}, both of which are calculated from the SED
\begin{eqnarray}
  T_\mathrm{bol} &=& 1.25 \times 10^{-11} \frac{\int_0^\infty \nu S_\nu \, d\nu}{\int_0^\infty S_\nu \, d\nu} \, \si{K} \nonumber \\
  \frac{L_\mathrm{smm}}{L_\mathrm{bol}} &=& \frac{\int_0^{c/\SI{350}{\micro\metre}} S_\nu \, d\nu}{\int_0^\infty S_\nu \, d\nu}. \nonumber
\end{eqnarray}
Table~\ref{tbl:evolutionary tracers} gives the definition of class boundaries for \LsubLbol and \Tbol.

In the standard picture of star formation \citep{Adams:1987gy}, a star is born in an isolated dense core, which is collapsing under its own gravity. The core eventually dissipates, revealing a pre-main-sequence star encircled by a massive disk. By applying the standard picture, the four observationally defined classes can be interpreted as an evolutionary sequence in which Class~0/I corresponds to a system still embedded within an infalling envelope, with Class~0 being those systems where more than half the mass still resides in the envelope \citep{Andre:1993cz}.

Although it is generally agreed that the progression through the observationally defined classes roughly correspond to a monotonic progression in time (e.g. \citealt{Evans:2009bk}), several authors (e.g. \citealt{Robitaille:2006cb,Dunham:2010bx}) have pointed out that there is not a one-to-one correspondence between the observationally defined classes and the physical evolution of protostellar systems. For example, in systems which contains a disk, \Tbol is known to be very sensitive to the orientation of the system relative to the observer. This led \citet{Robitaille:2006cb} to propose a distinction between observationally defined classes and physically defined stages. This distinction is followed here, where systems with $M_\star/M_\mathrm{env} < 1$ are referred to as Stage~0, and $M_\star/M_\mathrm{env} > 1$ as Stage~I. The mass within a radius of \SI{10000}{AU} from the protostar (diameter of \SI{0.1}{pc}) is used as a proxy of the envelope mass. In this section, we study the classification of protostars, focusing on the performance of \Tbol and \LsubLbol as evolutionary tracers of embedded protostars.

\subsection{Distribution of classes}
\label{sec:classdistribution}

\begin{table}
\caption{Definition of class boundaries following \citet{Andre:1993cz} and \citet{Chen:1995eo}. Note, that for \LsubLbol there is no boundary defined between Classes~I/II and~II/III.}             
\label{tbl:evolutionary tracers}      
\centering
\begin{tabular}{l c c }
\hline\hline
          & \LsubLbol & \Tbol \\
\hline
Class 0   & $\geq \SI{0.5}{\percent}$ & $< \SI{70}{\kelvin}$ \\
Class I   & $<    \SI{0.5}{\percent}$ & $\SI{70}{\kelvin} \leq T_\mathrm{bol} < \SI{650}{\kelvin}$ \\
Class II  & \ldots                    & $\SI{650}{\kelvin} \leq T_\mathrm{bol} < \SI{2800}{\kelvin}$ \\ 
Class III & \ldots                    & $\geq \SI{2800}{\kelvin}$ \\
\hline                  
\end{tabular}
\end{table}

\begin{figure*}
\centering
\includegraphics[width=18.6cm]{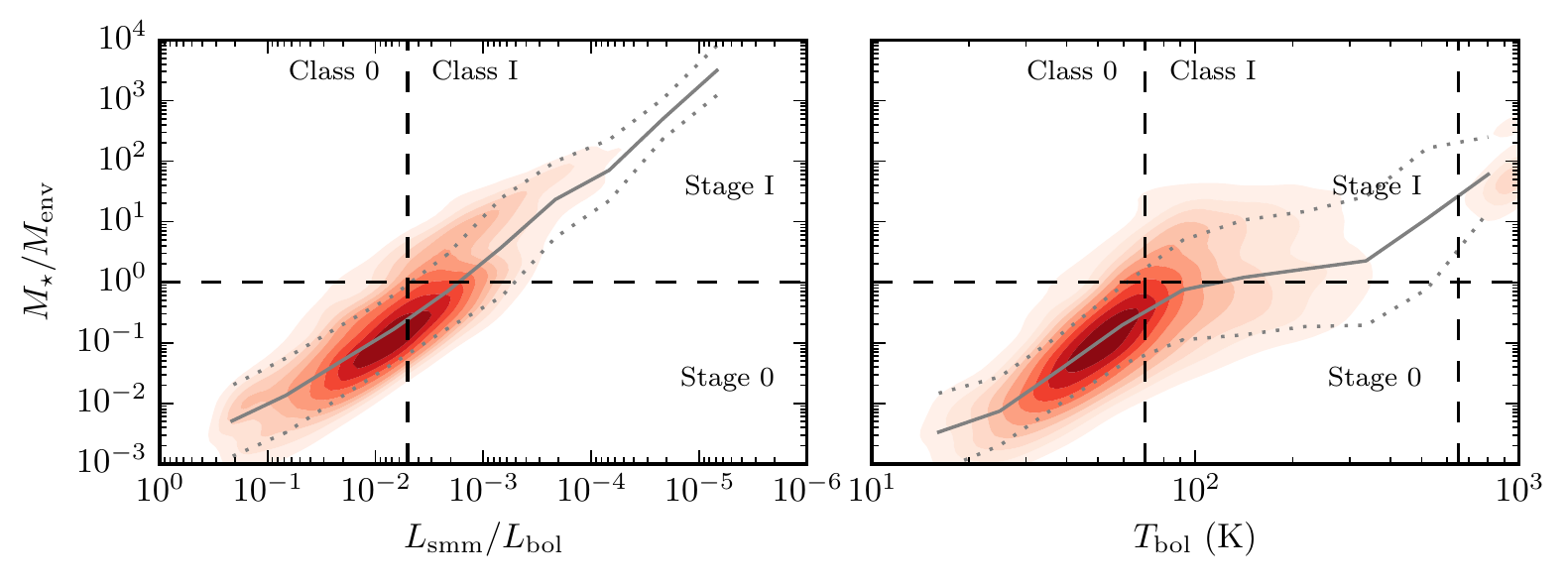}
   \caption{\LsubLbol and \Tbol vs.\ stage. The grey line is a binned median of the data, and the dotted lines indicate the one sigma uncertainties.}
      \label{fig:stagevstracers}
\end{figure*}

Figure~\ref{fig:TbolLsub} shows the distribution of \Tbol and \LsubLbol for the protostars in the simulation, along with real observations. The observations are a conjunction of data from c2d \citep{Evans:2009bk}, the Spitzer Gould Belt survey (GB) \citep{Dunham:2013do}, and the Herschel Orion protostar survey (HOPS) \citep{Fischer:2013jz,Manoj:2013ie,Stutz:2013ka}; also see \citet{Dunham:2014br}. The synthetic data in Fig.~\ref{fig:TbolLsub} are displayed as a probability density function. The contour levels are chosen so that the contours cover between \SI{90}{\percent} and \SI{10}{\percent} of the data in steps of \SI{10}{\percent}. This way of displaying the data, and the definition of the contour levels, is used throughout the paper.

Generally, the synthetic and observed distributions in Fig.~\ref{fig:TbolLsub} agree well with one another. The biggest difference between the two is that the fraction of protostars in the simulation with $T_\mathrm{bol} \gtrsim \SI{200}{\kelvin}$ is significantly reduced relative to the observations. This is a consequence of the spatial resolution in the simulation, which is \SI{8}{AU} in the best resolved regions. At high densities, such as those found very close to young protostars, this is not enough for the dust to become optically thin to radiation at wavelengths $\lesssim$\,\SI{100}{\micro\metre}. If the physical extent of the high-density region naturally cover several cells -- e.g. a deeply embedded Class~0 source, or the plane of a circumstellar disk -- this is no problem. However, in cases where the physical extent of the high-density region is smaller than one cell -- e.g. a disk viewed face-on -- it effectively makes the protostar look more embedded than it should be. For the types of objects and the wavelengths studied here, this is no concern. To study more evolved objects or shorter wavelengths, higher spatial resolution or a sub-scale model of the central cell around each protostar, is needed.

\subsection{Reliability of evolutionary tracers \Tbol and \LsubLbol}
\label{sec:reliability}

A central question, when dealing with evolutionary tracers, is how well they are able to predict the physical evolution of protostars. A natural first step in answering this question is to determine how often the two agree with each other. Based on the class boundaries, listed in Table~\ref{tbl:evolutionary tracers}, we find that \Tbol and \LsubLbol agree on the classification \SI{85}{\percent} of the time (see Table~\ref{tbl:classclass}). This is nearly equal to the \SI{84}{\percent} agreement reported by \citet{Dunham:2014br}.

\begin{table}
\caption{Fraction of synthetic data in each class quadrant in Fig.~\ref{fig:TbolLsub}.}
\label{tbl:classclass}
\centering
\begin{tabular}{l l c c c }
\hline\hline
         & & \multicolumn{3}{c}{\Tbol} \\
         & & Class II & Class I & Class 0 \\
\hline
\LsubLbol & Class 0 & \SI{0}{\percent} & \SI{7}{\percent}  & \SI{48}{\percent} \\
          & Class I & \SI{5}{\percent} & \SI{32}{\percent} & \SI{8}{\percent} \\
\hline
\end{tabular}
\end{table}

Figure~\ref{fig:stagevstracers} plots \Tbol and \LsubLbol vs.\ stage to study how well the evolutionary tracers agree with the physical stage. The figure shows that \LsubLbol and stage are tightly correlated throughout both the Class~0 and~I phases. \Tbol correlates well with stage during the deeply embedded Class~0 phase, while, in the less embedded Class~I phase, it does not. From Fig.~\ref{fig:stagevstracers} it can be seen that there is significant cross contamination, especially in the Stage~0/Class~I quadrant, however, this is easily explained as a consequence of the simplistic assumption made about the envelope masses. A more careful analysis of the actual masses of protostellar envelopes is beyond the scope of this work, and quantitative predictions about the relationship between the physically defined stages, and observationally defined classes, should therefore be avoided.

\begin{figure*}
\centering
\includegraphics[width=18.6cm]{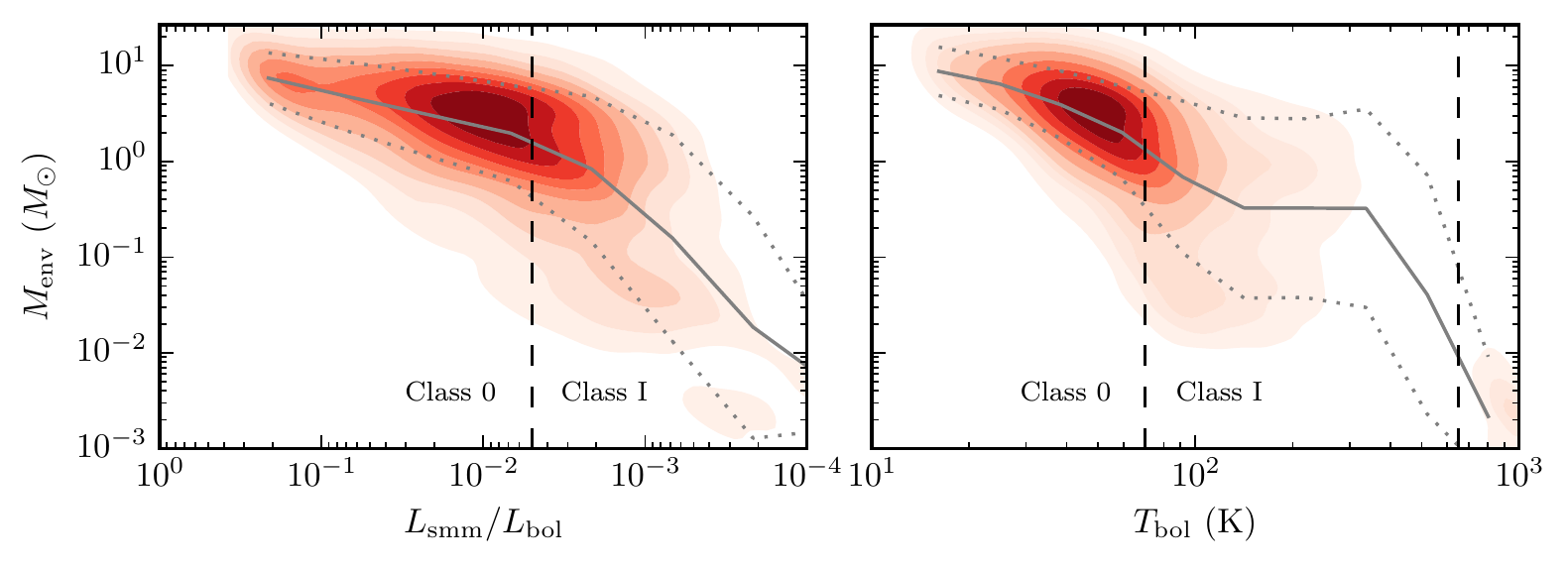}
   \caption{Envelope mass vs.\ \LsubLbol and \Tbol. The grey line is a binned median of the data, and the dotted lines indicate the one sigma uncertainties.}
      \label{fig:massvstracers}
\end{figure*}

\LsubLbol and \Tbol are designed to quantify the infrared excess of the SED, which depends on the amount of dust surrounding the protostar. Figure~\ref{fig:massvstracers} therefore shows envelope plotted mass against \LsubLbol and \Tbol. The figure shows that $M_\mathrm{env}$ does correlate with \LsubLbol and \Tbol, but that the correlation is not as strong as with the physical stage in Fig.~\ref{fig:stagevstracers}.

\begin{figure*}
\centering
\includegraphics[width=18.6cm]{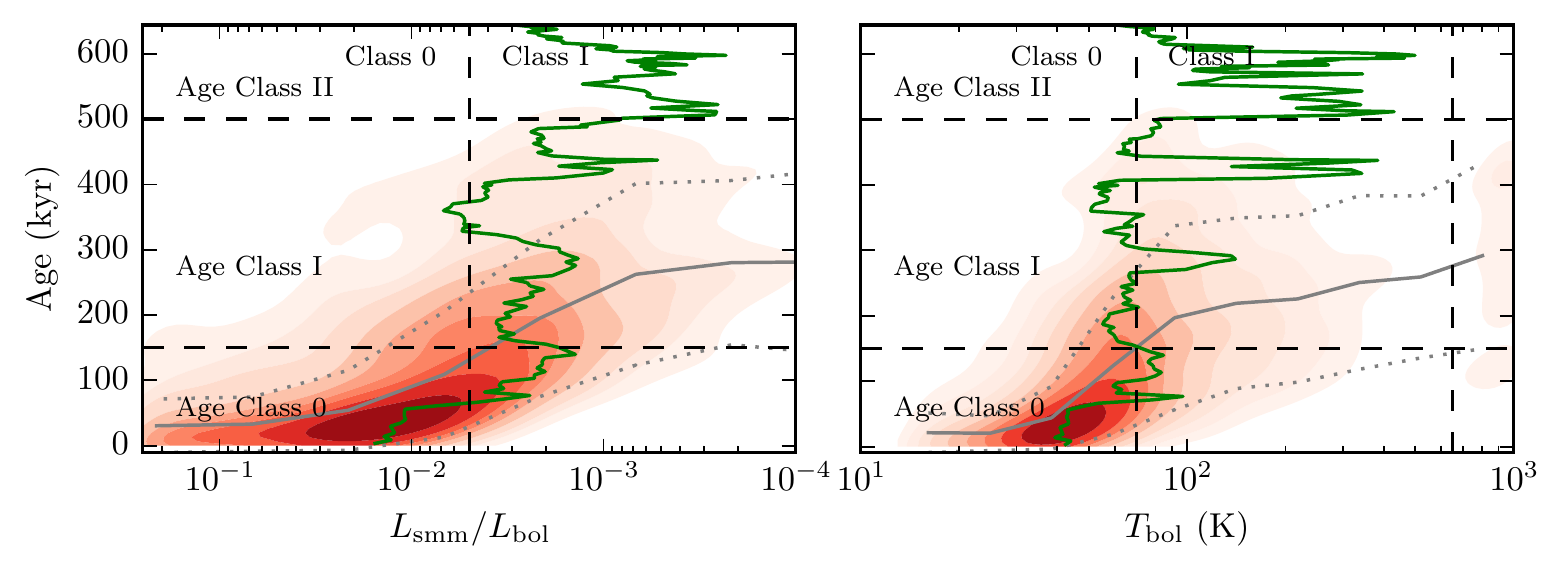}
   \caption{\LsubLbol and \Tbol vs.\ protostellar age. See Table~\ref{tbl:tracertime} for the fraction of points recorded in each quadrant. The green line shows the evolution of one protostar, which grows to a final mass of 3.6\,$M_\sun$. The grey line is a binned median of the data, and the dotted lines indicate the one sigma uncertainties.}
      \label{fig:tracertime}
\end{figure*}

\begin{table}
\caption{Fraction of synthetic data in each class quadrant in Fig.~\ref{fig:tracertime}.}             
\label{tbl:tracertime}
\centering
\begin{tabular}{l l c c c }
\hline\hline
         & & \multicolumn{3}{c}{\LsubLbol} \\
         & & Class 0 & Class I & Class II \\
\hline
          & Class II  & \SI{0}{\percent}  & \SI{2}{\percent}  & \ldots \\
Age       & Class I   & \SI{13}{\percent} & \SI{30}{\percent} & \ldots \\
          & Class 0   & \SI{42}{\percent} & \SI{13}{\percent}  & \ldots \\
\hline
         & & \multicolumn{3}{c}{\Tbol} \\
         & & Class 0 & Class I & Class II \\
\hline
          & Class II  & \SI{0}{\percent}  & \SI{1}{\percent}  & \SI{0}{\percent} \\
Age       & Class I   & \SI{14}{\percent} & \SI{25}{\percent} & \SI{4}{\percent} \\
          & Class 0   & \SI{42}{\percent} & \SI{13}{\percent} & \SI{1}{\percent} \\
\hline
\end{tabular}
\end{table}

Finally, it is instructive to study how \LsubLbol and \Tbol correlate with protostellar age. By counting the number of YSOs in each class, and assuming a Class~II lifetime of \SI{2}{Myr}, the approximate lifetimes of Class~0 and~I sources can be estimated to \SI{150}{kyr} and \SI{350}{kyr} respectively \citep{Evans:2009bk,Dunham:2014br}. The determination of accurate ages for young stars is very difficult \citep{Soderblom:2014be}, meaning that the assumed Class~II age may easily be wrong by a factor of two or more. At the same time, the fraction of YSOs in the different classes differs between individual clouds \citep{Evans:2009bk}, indicating that the local environments also play a role. Figure~\ref{fig:tracertime} shows protostellar age as function of \LsubLbol and \Tbol, and Table~\ref{tbl:tracertime} records the fraction of points in each quadrant. For ages <\,\SI{150}{\kilo\year}, Class~0 protostars (as measured by the SED) outnumber Class~I protostars by a factor 3.2 for both \LsubLbol and \Tbol. For ages between \SI{150}{\kilo\year} and \SI{500}{\kilo\year}, the Class~I/Class~0 ratio is roughly 2.3. The simulation has not been run for long enough to provide statistical information about systems older than \SI{500}{kyr}. The results do demonstrate an overall trend, in which older protostars are more likely to be less embedded and vice versa, but the scatter is substantial.

Figure~\ref{fig:tracertime} also includes an evolutionary track of an example protostar, which grows to a final mass of 3.6\,$M_\sun$. The track clearly shows that \LsubLbol and \Tbol are not monotonic functions of time, and that individual protostars may cross the Class~0/I boundary several times during their evolution. Generally, \LsubLbol and \Tbol are sensitive to changes in the total amount of dust surrounding the protostar, to changes in the distribution of dust in the system, and to changes of the protostellar luminosity. The influence of such effects are studied in the following section. The evolutionary track, shown in Fig.~\ref{fig:tracertime}, show that, at least for individual systems, \Tbol and \LsubLbol are poor indicators of age.

\begin{figure*}
\centering
\includegraphics[width=18.6cm]{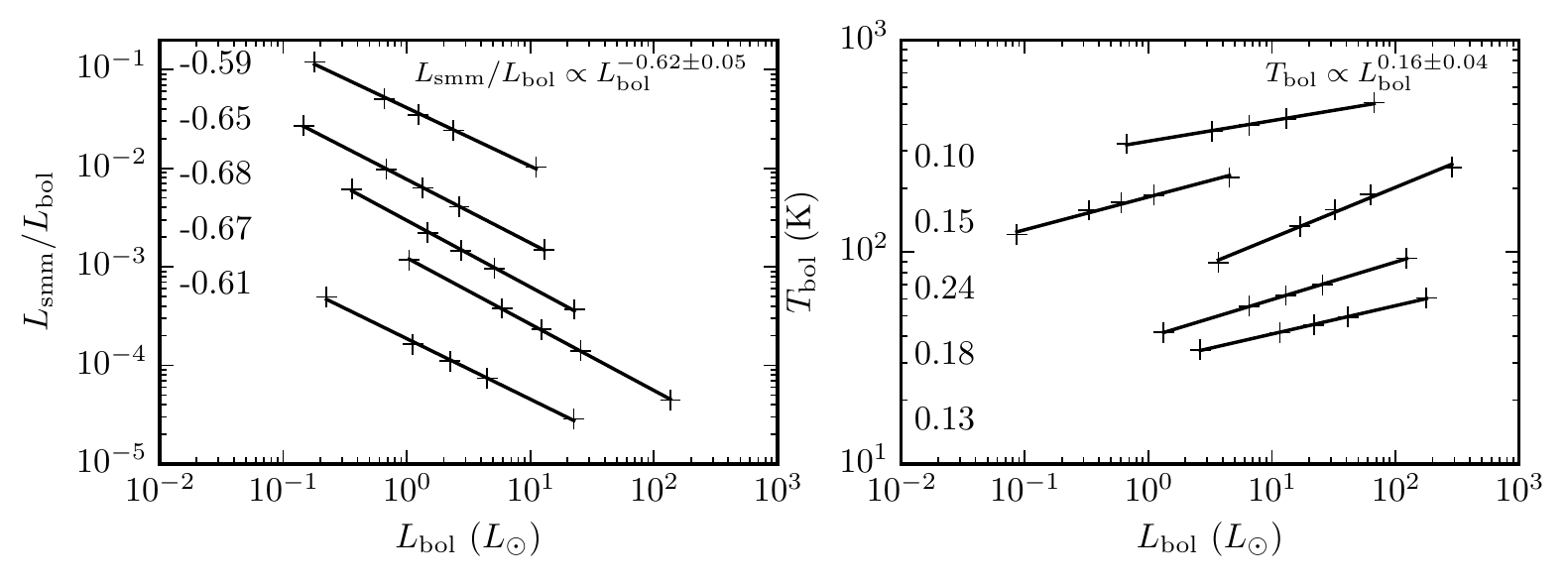}
   \caption{Dependence of \LsubLbol and \Tbol on $L_\mathrm{bol}$. The solid lines are power law fits to individual systems, and the numbers indicate the exponents of the fits. The results after fitting all 200 protostars in the sub-sample are shown in the upper right corner. The objects shown in the figure have been chosen to show the dependence on $L_\mathrm{bol}$ for a wide range of \LsubLbol and \Tbol values.}
      \label{fig:Lsuncomparison2}
\end{figure*}

Overall, the results of the section show that \LsubLbol and \Tbol agree on the classification most of the time, and that the marginal distributions of \Tbol and \LsubLbol match the observations well. Apart from the fact that \LsubLbol show a tighter correlation with physical stage throughout both the Class~0 and~I phases, relative to \Tbol, the analysis does not indicate that either tracer has a significant advantage over the other.

\subsection{Effective temperature, luminosity and projection effects}
\label{sec:luminosityandtemperature}

\Tbol and \LsubLbol are expected to depend on different parameters such as the effective temperature and luminosity of the central protostar, and the orientation of the system relative to the observer. In this section, the influence of these parameters on \Tbol and \LsubLbol are investigated.

So far, it has been assumed that all protostars in the simulation are perfect black bodies with temperatures of \SI{1000}{\kelvin}. This is clearly unrealistic; however, we find that changing the effective temperature of the central object, while keeping the luminosity constant, do not affect the results. This is because the emission from the protostar is completely reprocessed in the optically thick part of the envelope so that the exact shape of its spectrum becomes irrelevant.

The luminosity of the central protostar, on the other, hand does influence the shape of the observed SED. Luminous protostars heat their surroundings to high temperatures, and high temperature regions in disks and envelopes emit a larger fraction of their light at shorter wavelengths, making luminous protostars appear less embedded. We have extracted a sub-sample of 200 protostars, chosen at random from the original sample, and recalculated their SEDs after multiplying their luminosities by \numlist{0.1;0.5;2;10}. Changing the luminosity by a factor of two makes \SI{40}{\percent} and \SI{10}{\percent} of the protostars cross the Class~0/I boundary for \LsubLbol and \Tbol respectively. Changing the luminosity by a factor of ten changes these numbers to \SI{60}{\percent} and \SI{20}{\percent}. In Fig.~\ref{fig:Lsuncomparison2} the dependence of \LsubLbol and \Tbol on luminosity is illustrated for a few objects. The luminosity dependence is fitted well by a power law, and after fitting power laws to all protostars in the sub-sample we find
\begin{equation}
L_\mathrm{smm}/L_\mathrm{bol} \propto L_\mathrm{bol}^{-0.62 \pm 0.05} \quad \mathrm{and} \quad T_\mathrm{bol} \propto L_\mathrm{bol}^{0.16 \pm 0.04}. 
\label{eq:tracerluminosity}
\end{equation}
The results show that \LsubLbol is more sensitive to changes in the luminosity than \Tbol.

In systems with non-spherical geometry, the orientation of the system relative to the observer, will also affect \LsubLbol and \Tbol. \Tbol, in particular, is known to be very sensitive to projection effects, while \LsubLbol, which, contrary to \Tbol is unaffected by changes in the short wavelength emission, is expected to be less susceptible to changing projection. Probably the most extreme and simultaneously, one of the most common cases in which the orientation of a system relative to the observer is important for protostellar classification, is in case of the presence of a circumstellar disk. A disk viewed edge-on will appear more embedded than the same disk viewed face-on. Using the disk criterion adopted in Sect.~\ref{sec:diskinsimulations}, we have calculated edge- and face-on SEDs of more than \num{8000} disks in the simulation, to test if, and how much, \Tbol and \LsubLbol are affected. We find, when going from edge- to face-on, that \SI{30}{\percent} of the protostars change from Class~0 to~I for \LsubLbol, and \SI{50}{\percent} for \Tbol. Knowing that the systems in the simulation generally appear more embedded than they should for \Tbol (cf.\ Sect.~\ref{sec:classdistribution}), we expect this to be a lower limit. \LsubLbol changes its class roughly one time out of three, not because of $L_\mathrm{smm}$, which does not depend on the orientation of the system, but because of the measured luminosity, $L_\mathrm{bol}$, which, on average, increases by a factor \num{2.5} when going from edge- to face-on. This is a result of shielding by the dust, which means that a smaller fraction of the light escape through the disk's midplane relative to other directions.

\section{Comparing simulations and observations}
\label{sec:results}

\begin{figure*}
\centering
\includegraphics[width=18.6cm]{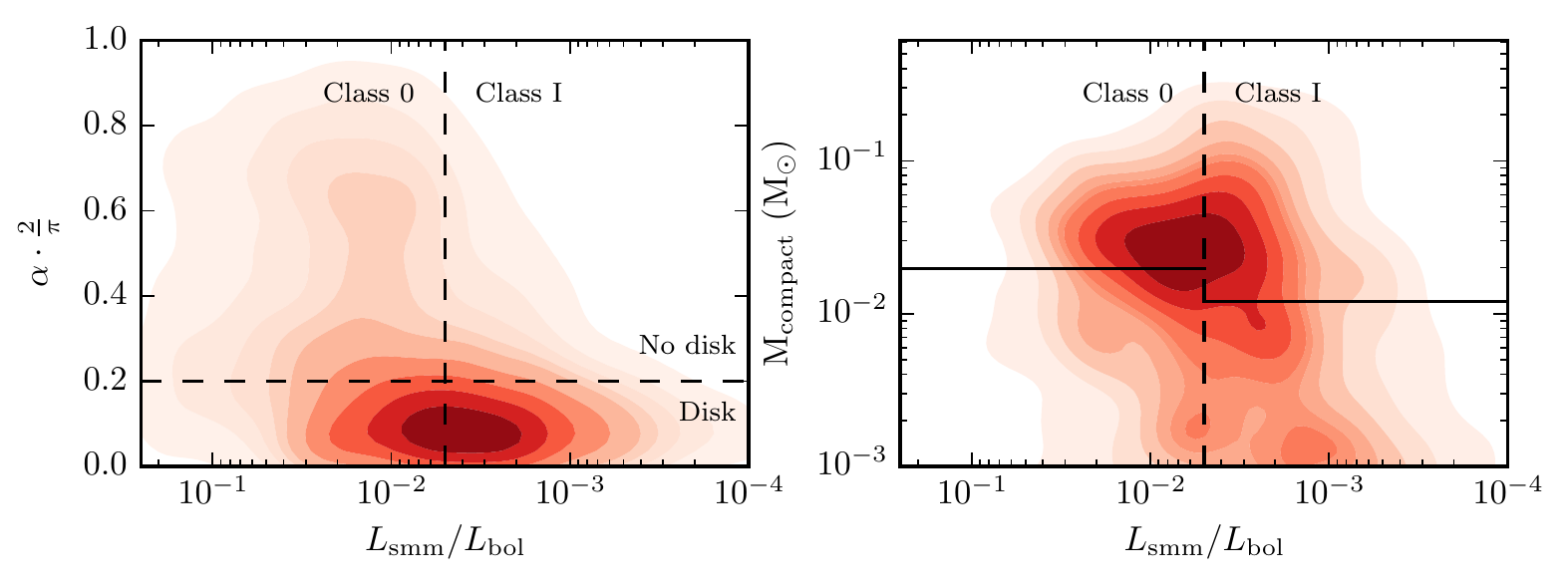}
\caption{Left: $\alpha$-angle vs.\ \LsubLbol. Right: compact mass vs.\ \LsubLbol. The right panel only includes systems with disks ($0 \leq \alpha \leq 0.2 \, \pi/2$). The compact mass may be regarded as an upper limit to the disk mass, since the contribution from the envelope has not been subtracted. The horizontal lines are median masses for the two classes.}
\label{fig:disk2}
\end{figure*}

\subsection{Disks and flux ratios}
\label{sec:disk}

The formation of circumstellar disks is a natural consequence of conservation of angular momentum during the core-collapse phase of star formation. Because of contamination from the envelope, direct detection of disks in embedded objects is very challenging, and requires high-resolution and high-sensitivity observations at long wavelengths. For this reason, there is still some uncertainty as to how early circumstellar disks actually form. In recent years, observational studies have demonstrated the presence of Keplerian disks around several Class~I protostars (e.g. \citealt{Brinch:2007gz,Lommen:2008em,Harsono:2014cu}), while for Class~0 protostars only three unambiguous detections have been reported so far \citep{Tobin:2012ee,Murillo:2013fn,Lindberg:2014bq}.

Equipped with the $\alpha$-angle method for disk detection, described in Sect.~\ref{sec:diskinsimulations}, we are able to answer fundamental questions about the properties of the disks in the simulation. The left panel of Fig.~\ref{fig:disk2} displays the angle $\alpha$ plotted against \LsubLbol, and shows that the majority (\SI{74}{\percent}) of the Class~I systems have values of $\alpha$ in the range $0 \leq \alpha \leq 0.2 \, \pi/2$, indicating the presence of a disk. This is in good agreement with observations, that report a disk fraction $\gtrsim$\,\SI{80}{\percent} in star forming regions younger than \SI{1}{Myr} \citep[e.g.][]{Wyatt:2008ht}. The disk fraction in the Class~0 objects is lower because many systems are still dominated by infall, but even so \SI{40}{\percent} still have values of $\alpha$ consistent with the presence of a disk.

The right panel of Fig.~\ref{fig:disk2}, which only includes systems that harbour a disk, shows compact masses vs.\ \LsubLbol. The ``compact mass'' is defined as the mass within a radius of \SI{400}{AU} from the protostar, and thus includes contributions from both the disk and the inner envelope. The compact mass can be regarded as an upper limit to the disk mass, and is used because of difficulties in disentangling disk and envelope masses.

\begin{figure*}
\centering
\includegraphics[width=18.6cm]{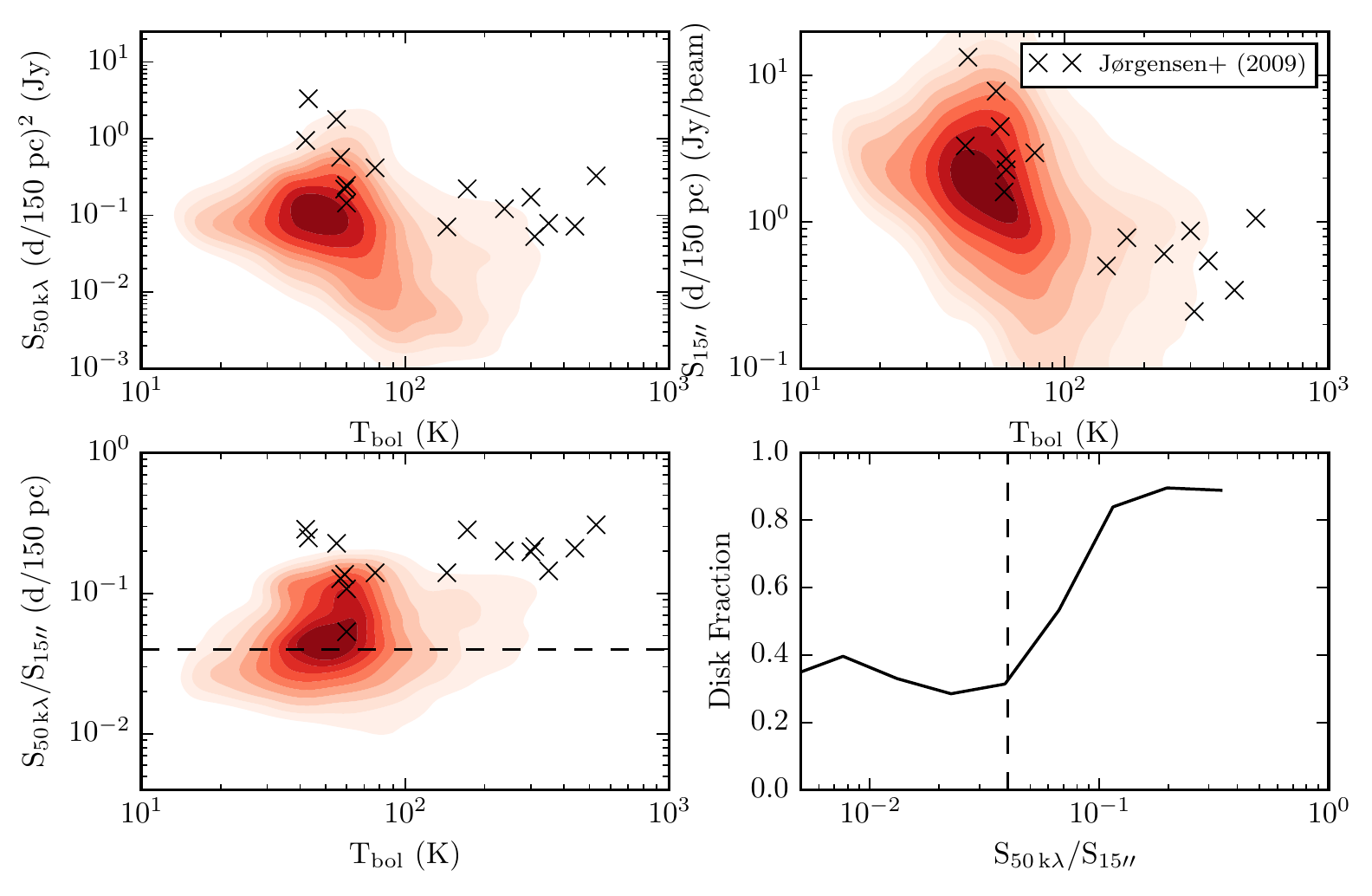}
   \caption{Top left: compact fluxes from the simulation compared directly to \SI{1.1}{mm} SMA observations. Top Right: extended fluxes from the simulation compared directly to \SI{0.85}{mm} SCUBA observations. The fluxes have been rescaled to a common distance of \SI{150}{pc}. For the interferometric fluxes this is just the inverse square law $F \propto d^{-2}$. For the single-dish fluxes we follow \citet{Jorgensen:2009bx} who, based on a density profile corresponding to a free-falling envelope ($\rho \propto r^{-1.5}$), found $F \propto d^{-1}$. Bottom left: flux ratios. The horizontal dashed line corresponds to the flux ratio expected for pure envelope emission \citep{Jorgensen:2009bx}. Bottom right: disk fraction vs.\ flux ratio. The solid line is the fraction of systems which contains a disk in each bin. The vertical dashed line is the same as the horizontal dashed line in the lower left panel.}
      \label{fig:fluxratio}
\end{figure*}

A number of studies have tried to disentangle the dust continuum emission between large-scale envelopes and circumstellar disks (e.g. \citealt{Looney:2003kz,Jorgensen:2005df,Eisner:2005jr,Lommen:2008em,Enoch:2011hu}). \citet{Jorgensen:2009bx} studied 10 Class~0 and 10 Class~I protostars, using a combination of interferometric and single-dish continuum observations, and developed a framework to interpret these observations based on comparisons with simple dust radiative transfer models. The single-dish observations, presented in \citet{Jorgensen:2009bx}, are from the James Clerk Maxwell Telescope (JCMT), have a wavelength of \SI{850}{\micro\metre}, a resolution of \ang{;;15}, and were used to measure the combined emission from the disk and envelope. The interferometric observations are from the SMA, have a wavelength of \SI{1100}{\micro\metre}, flux extracted at a baseline length of $50\, \mathrm{k}\uplambda$ (corresponding to a resolution of $\approx$\ang{;;4}), and were used to probe the disk emission, while resolving out the contribution from the envelope. Assuming optically thin dust, the compact interferometric flux, $S_\mathrm{50\, \mathrm{k}\uplambda}$, can be assumed to be directly proportional to the disk mass, while the extended single-dish flux, $S_\mathrm{15^{\prime\prime}}$, can be assumed to be proportional to the combined disk and envelope mass.

In recreating the continuum observations described above, we have assumed a source distance of \SI{150}{pc}, and rescale the observd fluxes of \citet{Jorgensen:2009bx} to match this distance. We have checked that the results presented in the following are not altered by changing the assumed distance to \SI{220}{pc}, which is the distance to most of the observed objects in \citet{Jorgensen:2009bx}. We assume a detection threshold of \SI{0.15}{Jy.beam^{-1}} for the synthetic single-dish observations \citep{Kirk:2006kt}, and \SI{10}{mJy} for the interferometric observations. \citet{Jorgensen:2009bx} converted their fluxes into disk and envelope masses, but, in order to make as few assumptions as possible, only the fluxes are compared here.

The top row of panels in Fig.~\ref{fig:fluxratio} shows compact and extended flux, $S_\mathrm{50\, \mathrm{k}\uplambda}$ and $S_\mathrm{15^{\prime\prime}}$, plotted against \Tbol. The synthetic compact fluxes are, on average, smaller by a factor of \num{\sim 6} relative to the observations; the synthetic extended fluxes are likewise, on average, smaller relative to the the observations by a factor of \num{\sim 2}. Adding the missing interferometric flux to the single-dish flux, brings it into agreement with the observations, showing that the reduced synthetic flux can be explained by the lack of compact emission alone. The lack of compact emission is likely due to the spatial resolution in the simulation not being sufficiently high to avoid numerical dissipation at the small spatial scales relevant for disks. Specifically, the numerical viscosity is artificially high on small scales leading to rapid accretion of material onto the protostar, which would have otherwise remained in the disk. This also means that the disk masses can be expected to be underestimated by the same factor.

The bottom left panel of Fig.~\ref{fig:fluxratio} shows the ratios between compact and extended fluxes. Assuming a spherical envelope model with $\rho \propto r^{-1.5}$, \citet{Jorgensen:2009bx} calculated that pure envelope emission is expected to yield a flux ratio $S_\mathrm{50\, \mathrm{k}\uplambda}/S_\mathrm{15^{\prime\prime}} = 0.04$, shown by the horizontal dashed line in the figure. A ratio above this value indicates the presence of an unresolved massive component, such as a disk. All but one of the systems presented in \citet{Jorgensen:2009bx} have a flux ratio consistent with the presence of a disk. Based on the discussion above, we expect the synthetic flux ratios to be smaller relative to the real ones with a factor \num{\sim 3}, which is also seen to be the case.

A central hypothesis of \citet{Jorgensen:2009bx} is that the flux ratio, between compact and extended emission, can be used as a tracer of disk occurrence. To test this hypothesis, the disk fraction has been plotted as function of flux ratio in the bottom right panel of Fig~\ref{fig:fluxratio}. The disk fraction is defined as the number systems containing a disk, divided by the total number of systems in each bin. The solid line in the figure is the disk fraction, and the dashed line indicates the limit of pure envelope emission. The disk fraction is roughly constant at \SI{\approx 30}{\percent}, with perhaps a shallow negative slope, up to a flux ratio of approximately \num{0.05}, above which is begins to climb rapidly. For flux ratios below \num{0.05} the total fraction of systems that contain a disk is \SI{33}{\percent}, while, for flux ratios above \num{0.05}, the fraction is \SI{76}{\percent}.

Because of the missing material on small scales, the flux ratios, measured from the synthetic observables, are systematically reduced relative to the observations. This precludes any quantitative comparison between observed and simulated flux ratios, since the uncertainties related to the small-scale physics in the simulation are considerable. Qualitatively, the results do demonstrate that a protostar is more likely to be encircled by a disk at higher flux ratios, supporting the hypothesis of \citet{Jorgensen:2009bx}.

\subsection{The protostellar luminosity function}
\label{sec:luminosities}

\begin{figure*}
\sidecaption
\includegraphics[width=12cm]{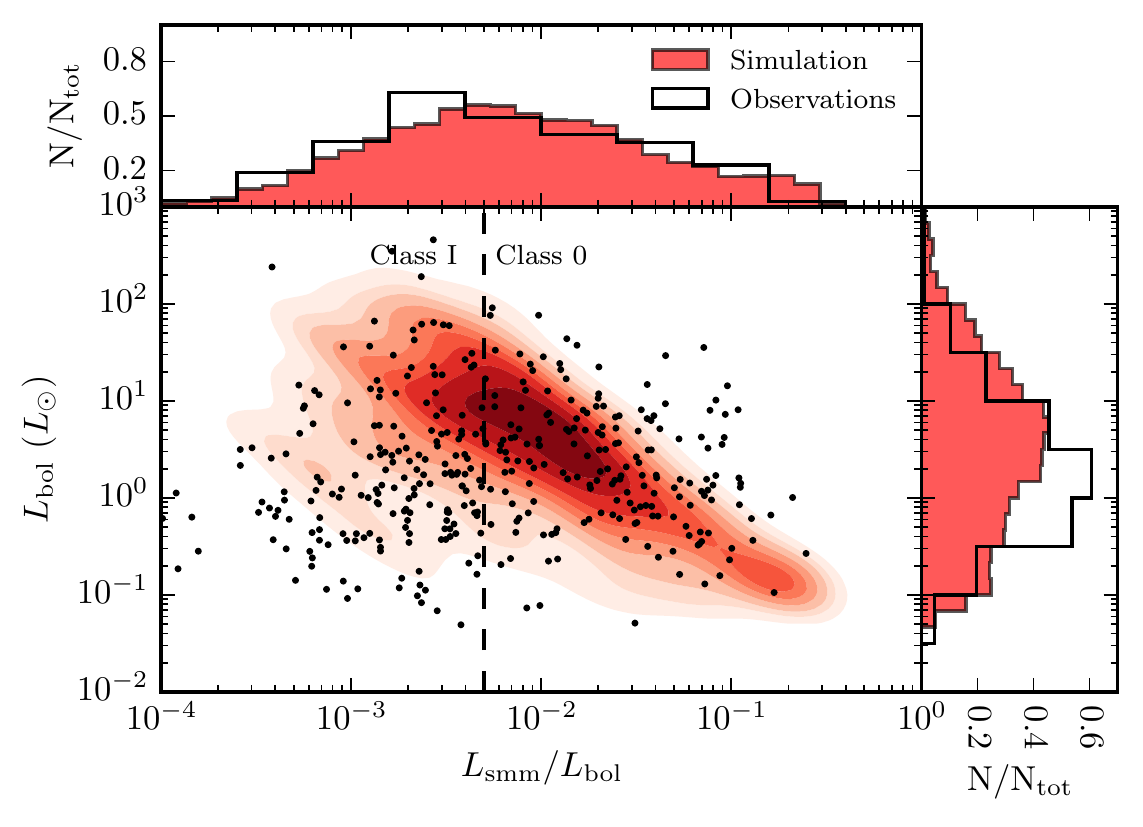}
\caption{Distribution of $L_\mathrm{bol}$ and \LsubLbol, for the protostars in the simulation (red contours), and observations (black dots). The marginal distributions of each variable are shown at the edges. The observational data is the same, as was used in Fig.~\ref{fig:TbolLsub}. Median luminosities of both simulation and observations are recorded in Table~\ref{tbl:luminosities}.}
\label{fig:LsmmLbolLacc}
\end{figure*}

The observed PLF is a roughly log-normal distribution, spanning more than three orders of magnitude, with a median luminosity of \num{\approx 1.3}\,$L_\sun$ \citep{Evans:2009bk,Dunham:2013do,Dunham:2014br}. A long-standing issue in low-mass star formation is the so-called ``luminosity problem'', where young stars are under-luminous with respect to expectations from simple physical models. The gravitational collapse of a spherical core, for example, yields an expected accretion rate, of \num{\sim e-5}\,$M_\sun \, \mathrm{yr}^{-1}$, which corresponds to $L_\mathrm{acc}$\,$\sim$\,30\,$L_\sun$, assuming a stellar mass of $0.25 \, M_\sun$ and radius of $2.5 \, R_\sun$; more than a factor of ten above the observed median.

The luminosity problem was first noticed by \citet{Kenyon:1990ki} who, as a possible solution suggested, that material accrete onto the protostar in short high-intensity bursts, giving rise to an episodic accretion paradigm. Observational evidence for episodic accretion include the FU~Orionis objects, which are pre-main-sequence stars undergoing accretion bursts raising their observed luminosity to \num{\sim 100}\,$L_\sun$ (see \citealt{Audard:2014gb} for a recent review). Recently, \citet{Jorgensen:2013ej} showed how water, left in the gas phase after an accretion burst, can explain the lack of HCO$^+$ around the deeply embedded protostar IRAS 15398–3359, thereby illustrating the potential of using chemical signatures as a tracer of episodic accretion.

Episodic accretion events, induced by disk instabilities, have been used with some success to reconcile models and observations (e.g. \citealt{Dunham:2012ic}). In a recent paper \citet{Padoan:2014ho} -- using a simulation similar to the one analysed here, but with lower spatial resolution, smaller box-size, and covering a longer time-span -- argued for a different paradigm in which accretion rates are regulated by turbulence induced variations in the large-scale mass infall from the envelope and onto the disk/star system.

\begin{table}
\caption{Median luminosities from observations and simulation. The first row is the observed luminosities of the combined c2d+GB+HOPS data \citep{Dunham:2014br}. The second row records the luminosities from the simulation.}
\label{tbl:luminosities}
\centering
\begin{tabular}{l c c c }
\hline\hline
                                         & Class 0 & Class I & All \\
                                         \hline
Observations                             &  1.9\,$L_\sun$    & 1.4\,$L_\sun$     & 1.7\,$L_\sun$ \\
Simulation                               &  2.1\,$L_\sun$    & 5.4\,$L_\sun$     & 3.3\,$L_\sun$ \\
\hline
\end{tabular}
\end{table}

Figure~\ref{fig:LsmmLbolLacc} plots $L_\mathrm{bol}$ against \LsubLbol in a fashion similar to the ``BLT'' diagrams first introduced by \citet{Myers:1993en}, but with \LsubLbol replacing \Tbol as an evolutionary tracer. Table~\ref{tbl:luminosities} records the median luminosities of the protostars in the simulation, as well as in the observations. The observational data are the same conjunction of c2d, GB, and HOPS data, that were used to study the distribution of \Tbol and \LsubLbol in Sect.~\ref{sec:classdistribution}. The simulated luminosities are, on average, a factor of two larger than the observed luminosities. However, the spread of the observed luminosities, spanning more than three orders of magnitude, is reproduced well by the simulation.

There is a natural difference between the inferred bolometric luminosity of any given source, measured by integrating over its SED, and its internal luminosity -- depending, for example, on the viewing angle towards sources where the surrounding dust is very asymmetrically distributed. Direct tests, comparing the bolometric and internal luminosities of the protostars in the simulation, reveal that the widths of their distributions are similar, with the median of the internal luminosity distribution being enhanced by a factor of \num{1.3} relative to the bolometric distribution.

The median luminosities of the Class~0 and~I protostars in the simulation are $2.1$\,$L_\sun$ and $5.4$\,$L_\sun$ respectively. For the Class~0 protostars this is close to the observed median of 1.9\,$L_\sun$, while, for the Class~I protostars, the simulated luminosities are enhanced by a factor \num{\sim 5} relative to the observations. Looking at Fig.~\ref{fig:LsmmLbolLacc}, observations and simulation agree well with one another, with the upper envelope of the simulated points following the relationship between \LsubLbol and $L_\mathrm{bol}$ given in Eq.~\eqref{eq:tracerluminosity}. The upper envelope of the real observations follow the same relationship except for \num{\sim 15} very embedded protostars at the Class~0 end, that are seen to fall above it. At the Class~I end of the figure the observations show a high density of observed systems at luminosities between \num{\sim 0.1}\,$L_\sun$ and~\num{\sim 5}\,$L_\sun$, that are not present to the same extent in the simulation. We believe this to be the objects that have evolved past their most embedded phase, and are lost in the simulation due to loss of spatial resolution (cf.\ discussion in Sect.~\ref{sec:postprocessing}). We also believe that the lack of these objects explain, why the median luminosity of the Class~I protostars in the simulation, as well as the sample as a whole, is larger than the observed values.

\subsection{Association between protostellar cores and protostars}
\label{sec:coreprotostar}

One of the fundamental question in star formation is how protostars accrete their mass. In the standard picture of low-mass star formation \citep{Adams:1987gy}, stars are born in dense molecular cloud cores, which also act as a mass reservoir for the protostars. In reality, most protostars are born in clusters, where dynamical interactions with other protostars may turn the process into a much more chaotic one. One of the questions in this area has been whether protostars stay in the dense environments, where they are born, throughout the main accretion phase, or if the situation is much more dynamic, where the motion of the protostars through the ambient medium, and the combined effect of the differential forces impacted by the turbulent ram pressure and the magnetic fields on the core compared to the protostar, is important for the accretion histories of protostars.

\begin{figure*}
\centering
\includegraphics[width=18.6cm]{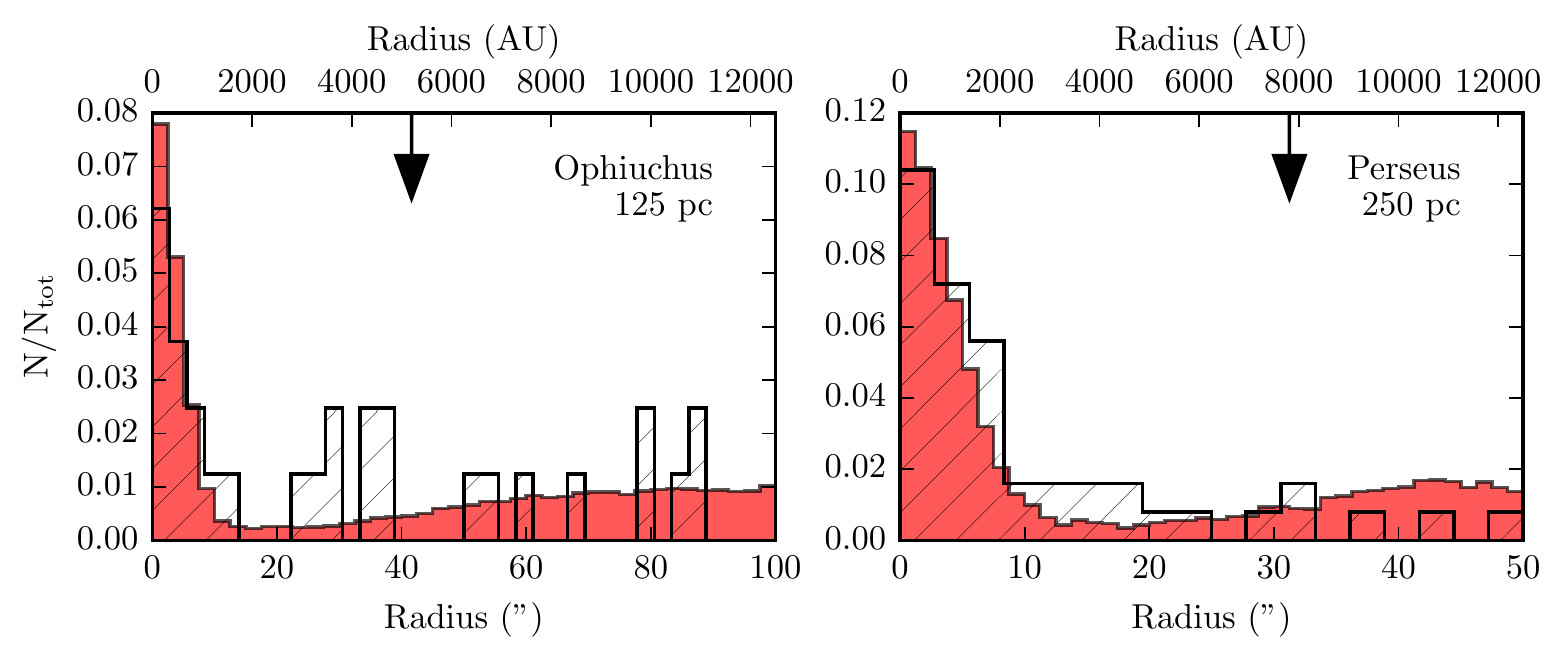}
\caption{Distribution of distances between cores and their nearest protostar in the simulation (shaded histograms), and from observations of Ophiuchus and Perseus (hatched histograms). The black arrows indicate the average core radius in the simulation.}
\label{fig:offsets}
\end{figure*}

The relationship between cores and protostars has been studied in several of the nearby molecular clouds \citep{Hartmann:2002ex,Jorgensen:2007ct,Jorgensen:2008gz,Enoch:2008hd}. These studies all conclude that, on average, the embedded protostars do not migrate far away from the dense cores where they were born. This finding stands in contrast to some numerical simulations, such as \citet{Bate:2012hy}, who found that the motion of protostars through the ambient medium plays a significant role for the accretion histories.

In this section, we study the protostellar cores in the simulation, with special focus on their association with embedded protostars. We compare our results to those of \citet{Jorgensen:2008gz}, who studied the properties of cores and protostars in the Ophiuchus and Perseus molecular clouds by utilising a combination of \SI{850}{\micro\metre} SCUBA continuum images and mid-infrared Spitzer data. To this end, we have created synthetic \SI{850}{\micro\metre} continuum images of all the protostars in the simulation, which are used to detect and characterise the protostellar cores. The method used for detecting cores in the simulation was described in Sect.~\ref{sec:cores}. \citet{Jorgensen:2008gz} used mid-infrared Spitzer observations to characterise and detect the positions of the observed protostars. These observations are not recreated, and the known positions of the protostars from the simulation are used instead. A normally distributed uncertainty, with FWHM of \ang{;;7}, is added to the positions of the protostars to emulate the uncertainty due to the size of the Spitzer beam, and pointing uncertainty in the submm observations. We follow \citet{Jorgensen:2008gz} and adopt a distance of \SI{125}{\parsec} to Ophiuchus, and \SI{250}{\parsec} to Perseus.

The number of detected cores varies depending on the assumed distance to the cloud. At a distance \SI{125}{\parsec}, we find a total of \num{177} cores in the simulation, while at \SI{250}{\parsec} we find \num{96} (last snapshot only). The decrease in the number of cores is partly a result of the less luminous cores no longer being detected at larger distances, partly due to the lower resolution of the maps, which serves to merge some cores. Nevertheless, many of the results presented below are independent on the assumed distance.

As discussed in Sect.~\ref{sec:generaloverview}, Class~0 and~I protostars are found to be closely associated with regions in the cloud of high column density, both in nature \citep{Evans:2009bk} and in the simulation. \citet{Jorgensen:2008gz} analysed the association between cores and protostars quantitatively by calculating the distance distribution between the two in Ophiuchus and Perseus. We repeat this analysis for the cores and protostars in the simulation, and plot the results in Fig.~\ref{fig:offsets}, which shows the distance distribution between cores and protostars in the simulation and in the observations of \citet{Jorgensen:2008gz}. The observed and synthesised distributions are seen to be very similar -- applying a two-sample Kolmogorov-Smirnov test yields p-values of \num{0.6} and \num{0.1} for Ophiuchus and Perseus respectively -- both of them peaking at small distances, confirming the close association between cores and protostars. 

The distance distribution is slightly different between Class~0 and~I protostars. Class~0 protostars are very narrowly distributed around the core centres, while the distribution of Class~I protostars is somewhat wider, although still centrally peaked. \SI{90}{\percent} of all Class~0 protostars lie within one core radius from their nearest core, while the same is true for \SI{70}{\percent} of the Class~I protostars. This is hardly surprising since Class~0 protostars are, by definition, deeply embedded objects, associated with high-density regions. For the distances corresponding to both Ophiuchus and Perseus, we find that \SI{\approx 60}{\percent} of the embedded protostars in the simulation lie within \ang{;;15} of the nearest core. In comparison, \citet{Jorgensen:2008gz} find that \SI{47}{\percent} of the embedded protostars in Ophiuchus and \SI{58}{\percent} in Perseus lie within \ang{;;15} of their nearest core, and simulation and observations are thus in good agreement with each other.

From Figs.~\ref{fig:offsets} and~\ref{fig:snapshot} we see that some cores are protostellar (contains a protostar) while others are starless. Assuming a distance of \SI{125}{pc} we find that \SI{16}{\percent} of the cores are protostellar, while, for a distance of \SI{250}{pc}, the fraction is \SI{24}{\percent}. Protostellar cores are, on average, more luminous than starless cores due to the presence of an internal source of luminosity. It is therefore also not surprising that the fraction increases with distance, since a lot of the starless cores are not luminous enough to be detected at the larger distance. For comparison, \citeauthor{Jorgensen:2008gz} found \SI{35}{\percent} the cores in Ophiuchus to be protostellar, and \SI{58}{\percent} in Perseus.

\begin{figure}
\includegraphics[width=\hsize]{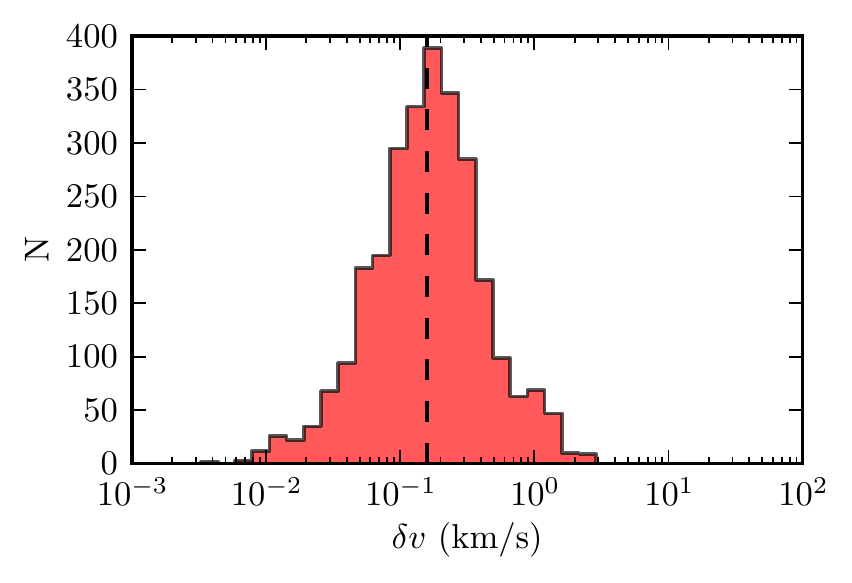}
\caption{Distribution of protostellar velocities relative to the gas and dust within a distance of \SI{5000}{AU} from the protostar. Only objects with no other protostars close by are included. The dashed line indicate the median value of the distribution.}
\label{fig:dv}
\end{figure}

Both \citet{Hartmann:2002ex} and \citet{Jorgensen:2007ct} used the close association found between cores and protostars to argue that the velocity dispersion of protostars relative to cores is very small. Based on the distribution of protostars around filaments, and assuming a stellar age of \SI{2}{Myr}, \citet{Hartmann:2002ex} estimated an upper limit on the velocity dispersion in Taurus of \SI{\approx 0.2}{\kilo\metre\per\second}. Using an analysis like the one shown in Fig.~\ref{fig:offsets}, and assuming protostellar age of \SI{0.1}{Myr}, \citet{Jorgensen:2007ct} estimated a velocity dispersion of \SI{\approx 0.1}{\kilo\metre\per\second} in Perseus. These estimates can be tested by measuring the two-dimensional velocity distribution, of the protostars in the simulation, relative to the dust and gas in their immediate vicinity. To avoid uncertainties due to dynamical interactions, we have only included embedded protostars with no other protostars within a radius of \SI{5000}{AU}. We also exclude protostars that have previously interacted dynamically with other protostars. The resulting distribution is shown in Fig.~\ref{fig:dv} and is seen to be roughly log-normal with a median velocity dispersion, $\delta v$, of \SI{\approx 0.15}{\kilo\metre\per\second}. \SI{60}{\percent} of the protostars have a $\delta v < \SI{0.2}{\kilo\metre\per\second}$, and \SI{90}{\percent} have $\delta v < \SI{0.5}{\kilo\metre\per\second}$. A manual inspection of the remaining protostars, with $\delta v > \SI{0.5}{\kilo\metre\per\second}$, reveals that most are either subjects to dynamical interactions with the large cluster seen in Fig.~\ref{fig:snapshot}, which is massive enough to interact with protostars even if no other protostars are present within \SI{5000}{AU}, or they are passing through regions where the gas has several velocity components.

\section{Summary}
\label{sec:summary}

This paper has presented an analysis of synthetic continuum images and SEDs, created from a large \SI{5 x 5 x 5}{\parsec} MHD simulation of a molecular cloud. Over the course of \SI{0.76}{Myr} the simulation forms more than \num{500} protostars, primarily within two sub-clusters. Having created more than \num{13000} unique radiative transfer models from the simulation, we have had access to an unprecedentedly large sample of synthetic observations, which have been compared to a number of observational studies. The main results of the paper are summarised as follows

\begin{enumerate}
  \item The simulation reproduces an extinction/column density threshold for cores, similar to that seen in observations (e.g. \citealt{Johnstone:2004ix,Enoch:2007dc}). Because the simulation is ideal MHD the threshold cannot be explained by the presence of photoionising radiation \citep{McKee:1989ho}. An alternative explanation is that the cores are a product of Jeans fragmentation and therefore primarily appear in the densest regions of the cloud \citep{Larson:1985to}.
  \item Values of the evolutionary tracers \Tbol and \LsubLbol are calculated for all the SEDs in the sample. We find that the agreement between observed and synthetic distributions of \Tbol and \LsubLbol is excellent, and that the two tracers agree on the classification of Class~0 and~I protostars \SI{85}{\percent} of the time, which is similar to the \SI{84}{\percent} agreement recorded from observations \citep{Dunham:2014br}. \LsubLbol correlates strongly with the physically defined stage over the entirety of its range. The same is true for \Tbol in the Class~0 phase, but not in the Class~I phase. Neither tracer correlates well with age, showing that \Tbol and \LsubLbol are poor indicators of this.
  \item Both \Tbol and \LsubLbol depend on parameters such as protostellar luminosity and the projection of the system relative to the observer. For individual sources the luminosity dependence is fitted well by a power law. We show that \LsubLbol is more sensitive to changes in the luminosity, while \Tbol, on the other hand, is more susceptible to projection effects.
  \item We devise a novel method for detecting disks in the simulation (the $\alpha$-angle method) based on the ratio between the radial and rotational motions of the gas around a protostar. This method, is found to be a simple, yet robust, way determining if a system contains a disk or not. Values of $\alpha$ lying within the range $0 \leq \alpha \leq 0.2 \, \pi/2$ are found empirically to indicate the presence of a disk. Power law fits to the rotation profiles show, that the disks found with the $\alpha$-angle method are consistent with Keplerian rotation \SI{80}{\percent} of the time. The remaining \SI{20}{\percent} are expected to be other kinds of rotationally dominated structures, for example, magnetically supported pseudodisks.
  \item Disks are found to form early on in the simulation, with one being found around \SI{40}{\percent} of the Class~0 protostars. For the Class~I protostars this fraction increases to \SI{74}{\percent}.
  \item Synthetic flux emission from the innermost regions around the protostars are found to be a factor of \num{\sim 6} too low relative to the observations of \citet{Jorgensen:2009bx}. The extended fluxes ares likewise found to be too small by a factor \num{\sim 2}. The missing flux is likely a result of numerical effects on the small spatial scales in the simulation, where high numerical viscosity may cause material, that would have otherwise remained in the disk, to accrete onto the protostar.
  \item \citet{Jorgensen:2009bx} used the flux ratio between compact and extended fluxes as an indicator of the presence of disks. In the simulation, we find that the disk fraction does increase with flux ratio; \SI{33}{\percent} of the systems in the simulation, with a flux ratio, $S_\mathrm{50\, \mathrm{k}\uplambda}/S_\mathrm{15^{\prime\prime}}$\,<\,0.05, contain a disk, while the fraction increases to \SI{76}{\percent} for ratios >0.05.
  \item The observed PLF is a wide distribution, spanning more than three orders of magnitude, with a median luminosity of \num{\approx 1.7}\,$L_\sun$. The bolometric luminosities, of the protostars in the simulation, reproduce the spread of the observed PLF, while the median is enhanced by a factor of two relative to observations. We believe the difference between the observed and simulated PLF is due to the simulated sample not being complete for Class~I sources.
  \item We find that protostars and cores are closely associated with one another, and that the distribution of distances between them is in excellent agreement with observations from Perseus and Ophiuchus. The relative velocity distribution between protostars, and the gas in their immediate surrounding, is roughly log-normal with a median of \SI{0.15}{km.s^{-1}}. Excluding dynamical interactions, protostars are, on average, not expected to migrate far away from the regions where they were born.
\end{enumerate}

Some weaknesses of the simulation have been illuminated during the work presented here. Most notably, the refinement criteria for the code should be adapted to depend on more parameters than local density, to make it possible to follow the protostellar evolution all the way to the Class~II phase. Other improvements, which will help obviating some of these weaknesses are the inclusion of sub-grid models for the inner-most cells, to deal with the resolution issue when doing radiative transfer, and stellar models for the protostars to accurately predict the luminosity.

Naturally, the ultimate goal is to understand the underlying physics of the star-formation processes, and thus also the assumptions in different flavours of simulations. The simulation includes the main ingredients to describe star formation in a piece of a molecular cloud: self-gravity, magnetic fields, driven turbulence, and sink particles. An important improvement will be to go beyond an isothermal equation of state, and to include ionising and non-ionising radiative feedback from the protostars, and cooling and heating processes. This is especially needed for a proper description of the interstellar medium near more massive stars. The spatial resolution of the simulation analysed in this paper is at the limits of what is currently computationally doable, but going to even higher physical resolutions, below \SI{1}{AU}, will be needed to account for some of the feedback from the protostars, and to resolve smaller disks around more evolved protostars.

With this paper, it has been demonstrated how direct comparison between observations and simulations is a very powerful tool, both in terms of interpreting observations, and in terms of testing different types of simulations to see how well they are able to reproduce observational results. This approach goes beyond comparing single selected objects with isolated models of star-forming cores and allow for statistical comparisons with observational surveys.

\begin{acknowledgements}

The authors acknowledge Michael Dunham and the HOPS team for providing the observational data used in Figs.~\ref{fig:TbolLsub} and~\ref{fig:LsmmLbolLacc}. We thank Christian Brinch, Michael Dunham, \AA ke Nordund, and Paolo Padoan for helpful discussions. We also thank the anonymous referee for providing helpful comments, which significantly improved the quality of the paper. This research was supported by a Lundbeck Foundation Group Leader Fellowship to JKJ as well as the European Research Council (ERC) under the European Union's Horizon 2020 research and innovation programme (grant agreement No 646908) through ERC Consolidator Grant ``S4F''. TH is supported by a Sapere Aude Starting Grant from the Danish Council for Independent Research. Research at Centre for Star and Planet Formation is funded by the Danish National Research Foundation. We acknowledge PRACE for awarding us access to the computing resource JUQUEEN based in Germany at Forschungszentrum J\"ulich for carrying out the simulation. Archival storage and computing nodes at the University of Copenhagen HPC center, funded with a research grant (VKR023406) from Villum Fonden, were used for the post-processing.

\end{acknowledgements}

\appendix
\section{Inclusion of multiple sources of luminosity}
\label{sec:multiplestar}

\begin{figure*}
\centering
\includegraphics[width=18.6cm]{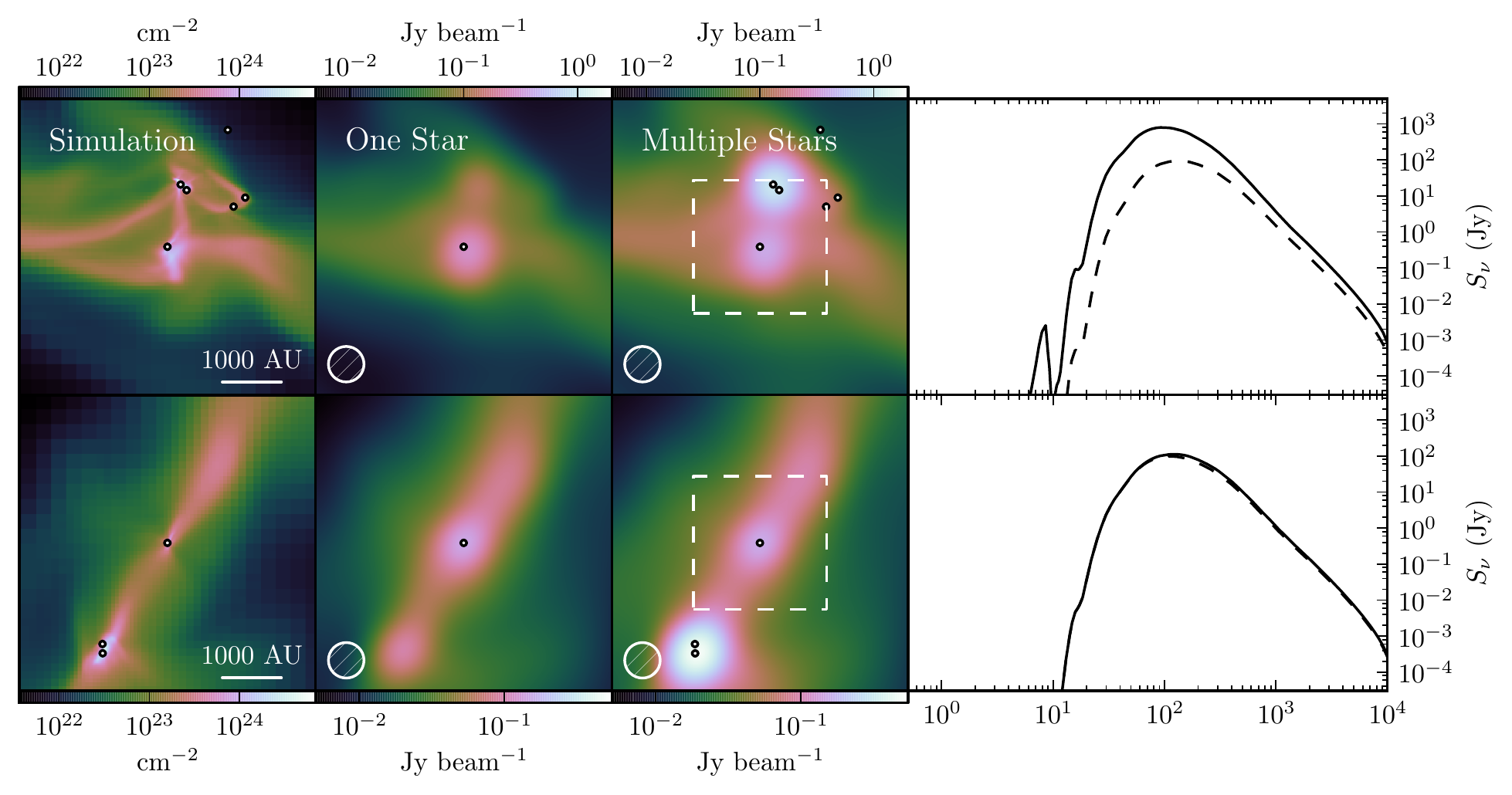}
   \caption{Two examples illustrating the effects of including multiple source of luminosity in the \radmc models. From left to right: projected gas column density, with dots indicating protostars; \SI{1100}{\micro\metre} \radmc continuum images with only one source of luminosity; same continuum image, but with multiple sources of luminosity; SEDs of the systems, where the single-luminosity SED is indicated by the dashed line and the multiple-luminosity SED by the solid line. We have assumed a distance of \SI{150}{pc} to the systems, and the continuum images have been convolved with a \ang{;;4} beam. The white dashed box in the images show the size and shape of the aperture over which the flux is integrated to calculate the SEDs. Selected physical parameters and observables of the central protostars in the two examples are presented in Table~\ref{tbl:proto2}.}
      \label{fig:continuumimage2}
\end{figure*}

\begin{table*}
\caption{Selected physical parameters and observables of the central protostars shown in Fig.~\ref{fig:continuumimage2}.}
\label{tbl:proto2}
\centering
\begin{tabular}{l c c c c c c c c c}     
\hline\hline
   & $M_\star$ & Age & $L_\mathrm{bol}$ & \Tbol & \LsubLbol & $S_\mathrm{50 \, k\uplambda, \, 1.1 \, mm}$ & $S_\mathrm{4^{\prime\prime}, \, 1.1 \, mm}$ & $S_\mathrm{15^{\prime\prime}, \, 0.85 \, mm}$ \\
   & ($M_\sun$)& (\si{kyr}) & ($L_\sun$) & (\si{K}) & (\si{\percent}) & (\si{mJy}) & (\si{mJy.beam^{-1}}) & (\si{Jy.beam^{-1}}) \\
\hline
\multicolumn{9}{c}{Top example from Fig.~\ref{fig:continuumimage2}} \\
\hline
One luminosity source       & 0.08 & 15.3 &  2.3 &  39 & 2.3  &  92 & 265 & 2.37 \\
Multiple luminosity sources &      &      & 24.5 &  51 & 0.7  & 550 & 377 & 5.49 \\
\hline
\multicolumn{9}{c}{Bottom example from Fig.~\ref{fig:continuumimage2}} \\
\hline
One luminosity source       & 0.25 & 43.2 &  2.8 &  46 &  1.3  &  53 & 150 & 1.60 \\
Multiple luminosity sources &      &      &  3.0 &  44 &  1.4  & 161 & 156 & 2.03 \\
\hline
\end{tabular}
\end{table*}

This appendix investigates the effects on the synthetic continuum images and SEDs of including multiple sources of luminosity in the \radmc models. The discussion is based on the two examples shown in Fig.~\ref{fig:continuumimage2}.

The first example, shown at the top of Fig.~\ref{fig:continuumimage2}, is a system consisting of six protostars. The central protostar has a luminosity of 2.3\,$L_\mathrm{bol}$ and an age of \SI{15.3}{kyr}. Four of the five remaining protostars in the cut-out have luminosities <\,0.2\,$L_\mathrm{bol}$, while the fifth protostar, which lies at a distance of \SI{1100}{AU} from the central protostar, has a luminosity of 29.7\,$L_\mathrm{bol}$. Table~\ref{tbl:proto2} records selected physical parameters and observables of the system, for both one and multiple sources of luminosity. The methods used for obtaining the observables have not been adjusted to take into account that there are more source of luminosity in the models, but are the same as used in the main paper. The luminous protostar falls inside the aperture used for calculating the SED, which is consequently affected significantly. This has adverse effects on both the measured luminosity as well as \LsubLbol. The measured fluxes, both interferometric and single-dish, are also affected by the addition of multiple luminosity sources. The interferometric flux, $S_\mathrm{50 \, k\uplambda, \, 1.1 \, mm}$, is particularly affected since the method used to extract this flux is not sensitive to the location of the flux emission, but only the magnitude. With only one source of luminosity per model this approach poses no problem, but naturally overestimates the flux when multiple sources are included. The single-dish fluxes are also affected -- the images in Fig.~\ref{fig:continuumimage2} have been convolved with a \ang{;;4} beam, which is seen to be a good enough resolution to separate the emission from central protostar from its more luminous countarpart. Still, $S_\mathrm{4^{\prime\prime}, \, 1.1 \, mm}$, increases with a factor of 1.4 when going from one to multiple sources of luminosity. For the larger beam, $S_\mathrm{15^{\prime\prime}, \, 0.85 \, mm}$, the situation is worse because the the two sources can no longer be separated.

The second example, shown at the bottom of Fig.~\ref{fig:continuumimage2}, is a system consisting of three protostars. The central protostar has a luminosity of 2.8\,$L_\sun$, and the most luminous of the two remaining stars, which is at a distance of \SI{2100}{AU} from the central protostar, has a luminosity of 7.9\,$L_\sun$. The final protostar in the system has a luminosity <\,0.1\,$L_\sun$. The observables in this example are somewhat less affected by the inclusion of multiple sources of luminosity, partly because the distance to the neighbouring protostars is larger relative to the first example, partly because the additional sources of luminosities are order of magnitude brighter than the central protostar. This also means that the SED and $S_\mathrm{4^{\prime\prime}, \, 1.1 \, mm}$ are unaffected, while the interferometric flux and $S_\mathrm{15^{\prime\prime}, \, 0.85 \, mm}$ continue to be affected.

The analysis of the two examples show that the effects of introducing more than one source of luminosity into the models can have quite adverse effects on the observables. When analysing real observations, it is typically possible to extract the signal from the source one is interested in, while filtering away the signal from other nearby sources. Such work is often done on an object-by-object basis and may include the application of custom apertures, flagging part of the data, and subtracting the signal one is not interested in. It is, in principle, possible to do the same for the synthetic observations, but it is not feasible due to the vast number of models, which is why we decided on doing a simple pipeline analysis, made possible by only including one source of luminosity per model.

\bibliographystyle{aa} 
\bibliography{citations} 

\end{document}